\documentclass[floatfix,prc,aps,twocolumn]{revtex4}
\usepackage{amssymb}
\usepackage{graphicx,psfrag}
\begin{document}

\newcommand{\note}[1]{\marginpar{\tiny {#1}}}   
\newcommand{\bld}[1]{\mbox{\boldmath $#1$}}     
\newcommand{\code}{$\mathtt{FREYA}$}
\newcommand{\Punf}{$^{239}$Pu(n$_{\rm th}$,f)}
\newcommand{\Pusf}{$^{240}$Pu(sf)}
\newcommand{\Cfsf}{$^{252}$Cf(sf)}
\newcommand{\Cmsf}{$^{244}$Cm(sf)}
\newcommand{\Unf}{$^{235}$U(n$_{\rm th}$,f)}
\newcommand{\Usf}{$^{238}$U(sf)}
\newcommand{\Punfnonf}{$^{239}$Pu}
\newcommand{\Pusfnosf}{$^{240}$Pu}
\newcommand{\Cfsfnosf}{$^{252}$Cf}
\newcommand{\Cmsfnosf}{$^{244}$Cm}
\newcommand{\Unfnonf}{$^{235}$U}
\newcommand{\Usfnosf}{$^{238}$U}
\newcommand{\beq}{\begin{eqnarray}}
\newcommand{\eeq}{\end{eqnarray}}
\newcommand{\SKIP}[1]{}
\newcommand{\nubar}{{\overline{\nu}}}
\newcommand{\half}{\mbox{$1\over2$}}
\title{Event-by-event study of neutron observables
in spontaneous and thermal fission}

\author{R.~Vogt$^{1,2}$ and J.~Randrup$^3$}

\affiliation{
$^1$Physics Division, Lawrence Livermore National Laboratory, Livermore, CA
94551, USA\break
$^2$Physics Department, University of California, Davis, CA 95616, USA\break
$^3$Nuclear Science Division, Lawrence Berkeley National Laboratory, 
Berkeley, CA 94720, USA}

\date{\today}

\begin{abstract}
The event-by-event fission model \code\ is extended
to spontaneous fission of actinides and a variety of neutron observables
are studied for spontaneous fission and fission induced by thermal neutrons
with a view towards possible applications for detection of
special nuclear materials.
\end{abstract}

\maketitle

\section{Introduction}

Phenomenological studies of nuclear fission are of particular interest for
possible practical applications in the fields of nonproliferation and security.
In particular, the detection of special nuclear material (SNM) has risen in
priority.  To better exploit all means of SNM detection, new efforts are 
underway to improve neutron detection technology, especially for the study of
fast fission neutrons from nuclear material.  Since all SNM emits neutrons,
it is advantageous to use these neutron emissions for the detection of such
material.  
For example, in highly enriched samples of plutonium (90\% $^{239}$Pu,
10\% $^{240}$Pu) and uranium  (90\% $^{235}$U, 10\% $^{238}$U), the small
content of $^{240}$Pu and $^{238}$U undergoes spontaneous fission, emitting
on average two neutrons per fission.  If it were possible to employ observable
differences in the characteristics of the fission process between the two
components of the material, it might be possible to distinguish between 
enriched and non-enriched samples of SNM.

Being of a penetrating nature, 
neutrons may provide specific signatures of SNM and thus have intrinsic
benefits over other observables.  Their long attentuation lengths imply
that they can propagate further through shielding material than photons emitted
by fission. Unfortunately their low cross sections in material constitute a
natural drawback to neutron-based detection methods: the longer time required
to obtain a clear measurement and the large solid angle necessary for detectors
to subtend to collect as many neutrons as possible \cite{AdamB}.  

The neutron background is relatively low, especially for prompt fission 
neutrons, emitted by the fragments directly after fission.  The dominant 
ambient neutron background to fission neutrons is from cosmic rays.
Neutron emission is thus
a fairly unique signature of fissile material.  Unfortunately, measurements
of the neutron energy spectra alone are not particularly useful for 
determining the isotopic content of a sample material.  
Interactions in matter, particularly through a shielding material, 
shifts the scattered neutrons to lower energy,
causing the yield of energetic fission neutrons to be reduced
by several orders of magntiude relative to the peak of the distribution.
Nevertheless, it is advantageous to study these neutrons using new 
fast-response detection systems
because fast neutrons preserve their direction better than thermal neutrons and
can thus be used in SNM detection schemes involving correlated observables
\cite{AdamB}.  
The present paper addresses such signatures of fission.

Heretofore, most fission simulations have assumed 
that all emitted neutrons are drawn from the same energy spectrum 
which precludes correlations between the neutron mutliplicity 
and the associated spectral shape.
In our event-by-event treatment, 
such inherent correlations are automatically included
and we examine them with an eye toward specific applications.  
Our approach employs the fission model \code\
(Fission Reaction Event Yield Algorithm)
which incorporates the relevant physics 
with a few key parameters determined 
by comparison to data \cite{VRPY,RV,VRBDO}.
It simulates the entire fission process and produces
complete fission events with full kinematic information on the emerging
fission products and the emitted neutrons and photons, incorporating 
sequential neutron evaporation from the fission fragments.  (We will examine
prompt fission photon
production in a later publication.)
\code\ provides a means of using readily-measured observables 
to improve our understanding of the fission process and
it is, therefore, a potentially powerful tool for bridging the gap 
between current microscopic models and important fission observables 
and for improving estimates of the fission characteristics 
important for applications.  

We compare and contrast correlations between neutron observables in
spontaneous and thermal fission of $^{240}$Pu,
$^{240}$Pu(sf) and $^{239}$Pu(n$_{\rm th}$,f) respectively,
as well as between $^{238}$U(sf) and $^{235}$U(n$_{\rm th}$,f).
We also study these observables in the spontaneous
fission of $^{252}$Cf, often
used as a calibrator for other fission measurements, and $^{244}$Cm.

In the next section, we describe the experimental data employed,
in particular the fission fragment mass distributions and 
the total fragment kinetic energy as a function of fragment mass.  
We then discuss various neutron observables, including 
the prompt fission neutron multiplicity as function of fragment mass, 
the neutron multiplicity distribution,
and the energy spectrum of the prompt fission neutrons.
We also study the neutron-neutron angular correlations as well as 
the correlations between both
the total kinetic energy of the fission products and 
their residual excitation energy 
as a function of the total neutron multiplicity.  
We finally discuss potential ways to exploit these correlations 
and conclude with some final remarks.

\section{Mass and charge partition}
\label{split}

The treatment of spontaneous fission in \code\ is similar to that of 
neutron-induced fission, except for the simplification that there is 
no possibility for any pre-fission emission.
Thus, generally, we start with a fissile nucleus $^{A_0}Z_0$ 
having a specified excitation energy $E_0^*$,
and let it undergo binary fission into a heavy fragment $^{A_H}Z_H$ 
and a complementary light fragment $^{A_L}Z_L$.
The fragment masses are obtained from experimental mass yields
by the procedure employed in the original description of \code\ \cite{RV}.

The fragment mass yields, $Y(A)$, are assumed to exhibit 
three distinct modes of Gaussian form \cite{g-fit},
\begin{equation}\label{eq:defs1s2}
Y(A)\ =\ S_1(A)+S_2(A)+S_L(A)\ .
\end{equation}
The first two terms represent asymmetric fission modes
associated with the spherical shell closure at $N=82$
and the deformed shell closure at $N=88$, respectively,
while the last term represents a broad symmetric mode.
The symmetric mode is relatively insignificant 
for spontaneous fission which is at rather low nuclear excitation.  The
exception is $^{252}$Cf with a comparatively large symmetric contribution.

The asymmetric modes have a two-Gaussian form,
\begin{eqnarray}
S_i = {{N}_i\over\sqrt{2\pi}\sigma_i}\left[
 {\rm e}^{-(A-\bar{A}-D_i)^2/2\sigma_i^2}\!
+{\rm e}^{-(A-\bar{A}+D_i)^2/2\sigma_i^2}
\right]\! ,
\label{s1s2}
\end{eqnarray}
while the symmetric mode is given by a single Gaussian
\begin{eqnarray}
S_L = {N_L\over\sqrt{2\pi}\sigma_L}\,
	{\rm e}^{-(A -\bar{A})^2/2\sigma_L^2}\ ,
\label{slong}
\end{eqnarray}
with $\bar{A}=\mbox{$1\over2$}A_0$.
Since each event leads to two fragments, the yields are normalized so that 
$\sum_AY(A)=2$.  Thus,
\begin{eqnarray} 
2{N}_1+2{N}_2+{N}_L=2 \, \, ,
\label{nsum}
\end{eqnarray}
apart from a negligible correction 
because $A$ is discrete and bounded from both below and above.

\begin{figure}[tbp]
\includegraphics[angle=0,width=\columnwidth]{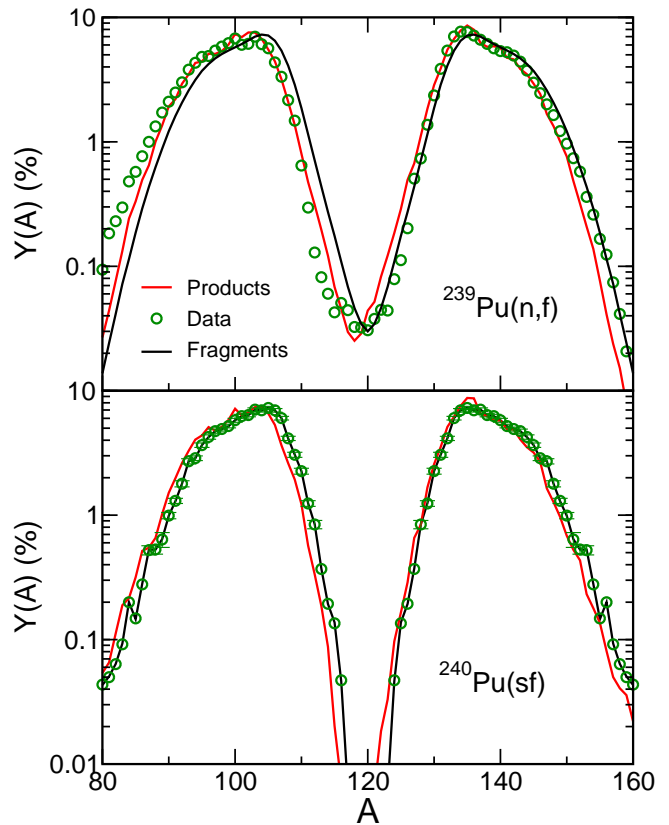}
\caption[]{(Color online) The percent yield as a function of fragment mass
for $^{239}$Pu(n$_{\rm th}$,f) \protect\cite{EnglandRider} (top) and 
$^{240}$Pu(sf) \protect\cite{SchillebeeckxPusf} (bottom).  The product yield
data are shown for \Punf\ while the fragment yields are given for \Pusf. 
The black curves are the 5-Gaussian fits to the fragment distributions while
the  red curves are the results after neutron emission in \code.}
\label{yields_Pu_fig}
\end{figure}

\begin{figure}[tbp]
\includegraphics[angle=0,width=\columnwidth]{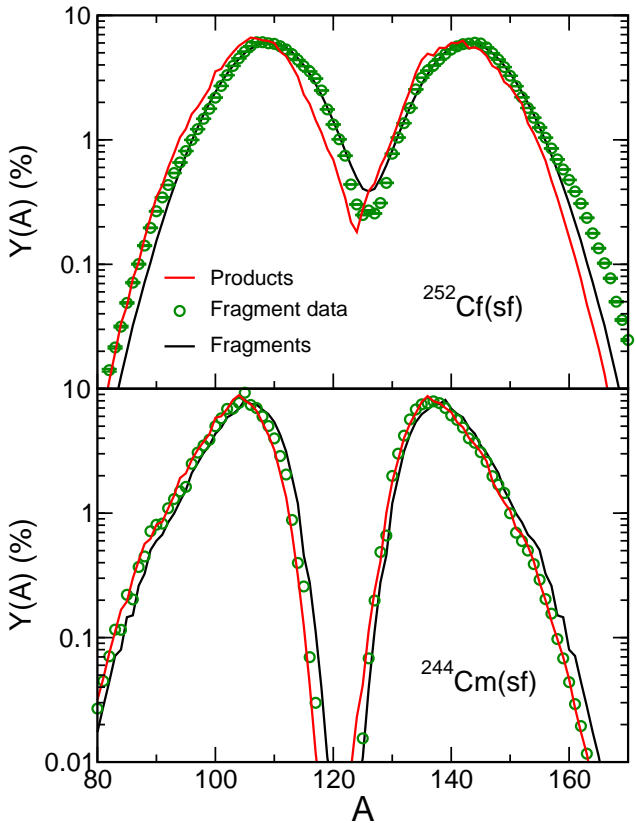}
\caption[]{(Color online) The percent yield as a function of fragment mass
for $^{252}$Cf(sf) \protect\cite{HambschCf} (top) and 
$^{244}$Cm(sf) \protect\cite{SchmidtCmTKE} (bottom).  The data are fission 
fragment measurements.
The black curves are the 5-Gaussian fits to the fragment distributions while
the  red curves are the results after neutron emission in \code.}
\label{yields_CfCm_fig}
\end{figure}

\begin{figure}[tbp]
\includegraphics[angle=0,width=\columnwidth]{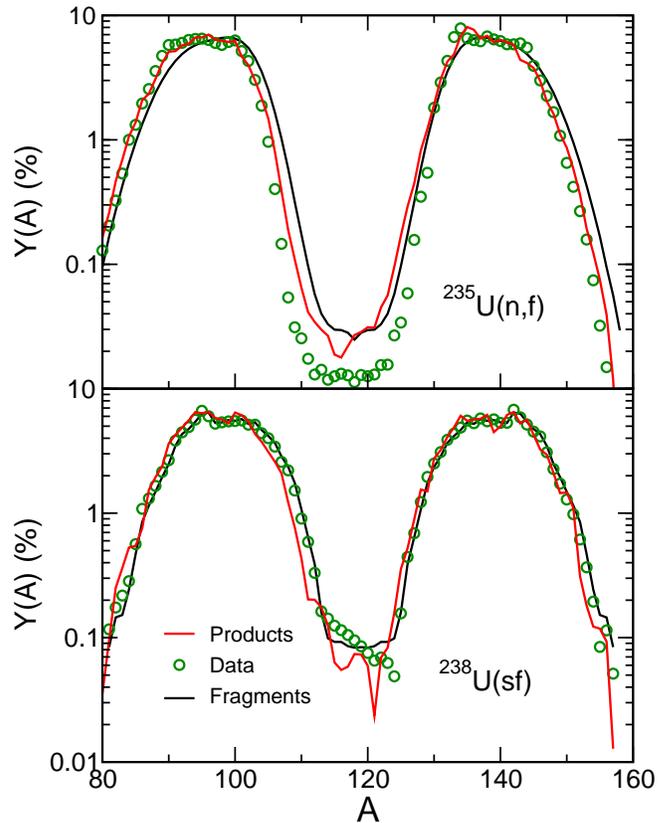}
\caption[]{(Color online) The percent yield as a function of fragment mass
for $^{235}$U(n$_{\rm th}$,f) \protect\cite{EnglandRider} (top) and 
$^{238}$U(sf) \protect\cite{IvanovUsf} (bottom).  The product yield
data are shown for \Unf\ while the fragment yields are given for \Usf. 
The black curves are the 5-Gaussian fits to the fragment distributions while
the  red curves are the results after neutron emission in \code.}
\label{yields_U_fig}
\end{figure}

The results are shown for the fission fragment and the subsequent product
yields in Figs.~\ref{yields_Pu_fig}-\ref{yields_U_fig}.  
The yields reported for spontaneous fission
are for the (primary) fragments while those reported for inducd fission
are for the (post-evaporation) products \cite{EnglandRider}.
The modeling of the fission fragment yields for neutron-induced fission over a
range of incident neutron energies is
discussed in Ref.~\cite{VRBDO}.  The product yields are obtained after \code\
has finished emitting neutrons from the excited fragments.  All the yields 
exhibit similar behavior, a rather broad double-humped distribution with a
gap near symmetry, $A_0/2$.  The symmetric contribution is typically very 
small.

The results in Fig.~\ref{yields_Pu_fig} are most closely related because
both $^{239}$Pu(n$_{\rm th}$,f) and $^{240}$Pu(sf) start from a compound 
nucleus with the same value of $A_0$.
The \Pusf\ data were taken from a study of $^{238,240,242}$Pu(sf) relative to 
$^{239}$Pu(n$_{\rm th}$,f) \cite{SchillebeeckxPusf}.  
The experiment was set up next to a reactor so that $^{239}$Pu(n$_{\rm th}$,f)
could be used as a calibrator,
with a large acceptance geometry to partially compensate 
for the low rate of spontaneous fission.  
(The highest collected total number of spontaneous fission events,
about 12000 for \Pusf, 
was a factor of about 200 below the number of thermal neutron-induced events.) 
The somewhat larger widths of the mass distributions resulting from
thermal neutron-induced fission were attributed to increased 
intrinsic excitation energy near the scission point \cite{SchillebeeckxPusf}.  

The \Cfsf\ fragment yields, 
shown in the upper panel of Fig.~\ref{yields_CfCm_fig},
result from an analysis of $2.5 \times 10^8$ events \cite{HambschCf}.  The
experiment focused on the far asymmetric mass region and showed that the
enhancements in the yields observed previously were due to the choice of
angular selection criteria.  Choosing $\cos \theta > 0.9$ eliminated events
where energy loss in the foil is large.  We choose these results for use in
\code, even though our focus is not on the far-asymmetric region, because the 
large sample size provides more accurate input.  
The \Cfsf\ yields were also measured with
\Cmsf\ in Ref.~\cite{SchmidtCmTKE}, albeit with fewer statistics, $5.4 \times
10^5$ \Cfsfnosf\ fissions and 71000 \Cmsfnosf\ fissions.  The \Cmsfnosf\ 
fragment mass distributions and average fragment masses reported in 
Ref.~\cite{SchmidtCmTKE} are in good agreement with previous measurements, thus
we can accept their reported yields with some confidence.

The \Usf\ yield data in Fig.~\ref{yields_U_fig} was obtained from a 
uranium sample with a natural isotopic composition, {\em i.e.} with a small
admixture of \Unfnonf\ \cite{IvanovUsf}.  A \Cfsf\ neutron source was placed 
outside a double ionization chamber to provide thermal neutrons and thus
allow comparison of the fission characteristics of \Usf\ with \Unf.  
The double ionization
chamber allowed measurements of the yields and kinetic energies of the two
fission fragments in coincidence.  They corrected for energy loss in the 
backing material and excluded angles greater than 60$^\circ$ to eliminate
events where the fragments passed through more material, necessitating a larger
correction for energy loss in matter.  The results shown here are based on
2800 fission events.  They noted more fine structure in \Usf\ than in thermal
neutron-induced fission of \Unfnonf.  The yield at $A \sim 119$ is very poorly
determined \cite{IvanovUsf}.  Unfortunately, no other \Usf\ data were found
for comparison.

In all cases, the
locations of the asymmetric peaks in the data are similar while there appears
to be a clearer separation of the asymmetric peaks at symmetry for the 
spontaneously fissioning nuclei.  The asymmetric Gaussians also appear to be
somewhat narrower in the case of spontaneous fission with the exception of
\Cfsfnosf, as shown in Fig.~\ref{yields_CfCm_fig}.  
In this case, the tails of the asymmetric distributions shown
are quite broad so the dip at symmetry is filled in to a considerable degree,
even relative to neutron-induced fission.

A clear shift between the fragment yields (before neutron emission) and
the product yields (after neutron emission) is apparent in all cases.  
The magnitude of the shift depends on the overall mean neutron multiplicity, 
$\nubar$, which in turn depends on the partition of the excitation energy 
between the light and heavy fragments.  
The shift is not symmetric but is larger for the light fragment, 
especially near symmetry.  The location of
the peak in the heavy fragment yield at $A \approx 130$ does not exhibit a 
significant shift due to neutron emission in any of the cases shown, even 
though the shift is apparent for other values of $A$.  This is
due to the proximity of the doubly-magic closed shell with $Z_H = 50$ and 
$N_H = 82$.  This behavior is also apparent in the shape of TKE($A_H$) 
and in the dependence of the mean neutron multiplicity on fragment mass,
$\nubar(A)$, as will be discussed later.

The fragment charge, $Z_f$, is selected subsequently.
For this we follow Ref.~\cite{ReisdorfNPA177} and employ a Gaussian form,
\begin{equation}
P_{A_f}(Z_f)\ \propto\ {\rm e}^{-(Z_f-\bar{Z}_f(A_f))^2/2\sigma_Z^2}\ ,
\end{equation}
with the condition that $|Z_f-\bar{Z}_f(A_f)|\leq5\sigma_Z$.
The centroid is determined by requiring that the fragments have, on average,
the same charge-to-mass ratio as the fissioning nucleus,
$\bar{Z}_f(A_f)=A_fZ_0/A_0$.  The dispersion is 
the measured value, $\sigma_Z=0.5$ \cite{ReisdorfNPA177}.
The charge of the complementary fragment then follows using $Z_L+Z_H=Z_0$.

\section{Fragment energies}

Once the partition of the total mass and charge among the two fragments
has been selected, 
the $Q$ value associated with that particular fission channel follows 
as the  difference between the total mass of the fissioning nucleus
and the ground-state masses of the two fragments,
\begin{equation}
Q_{LH}\ =\ M(A_0) - M_L - M_H\ .
\end{equation}
\code\ takes the required nuclear ground-state masses
from the compilation by Audi {\em et al.}~\cite{Audi},
supplemented by the calculated masses of M{\"o}ller {\em et al.}~\cite{MNMS}
when no data are available.
The $Q_{LH}$ value for the selected fission channel is then divided up between
the total kinetic energy (TKE) and the total excitation energy (TXE)
of the two fragments.

\begin{figure}[tbp]
\includegraphics[angle=0,width=\columnwidth]{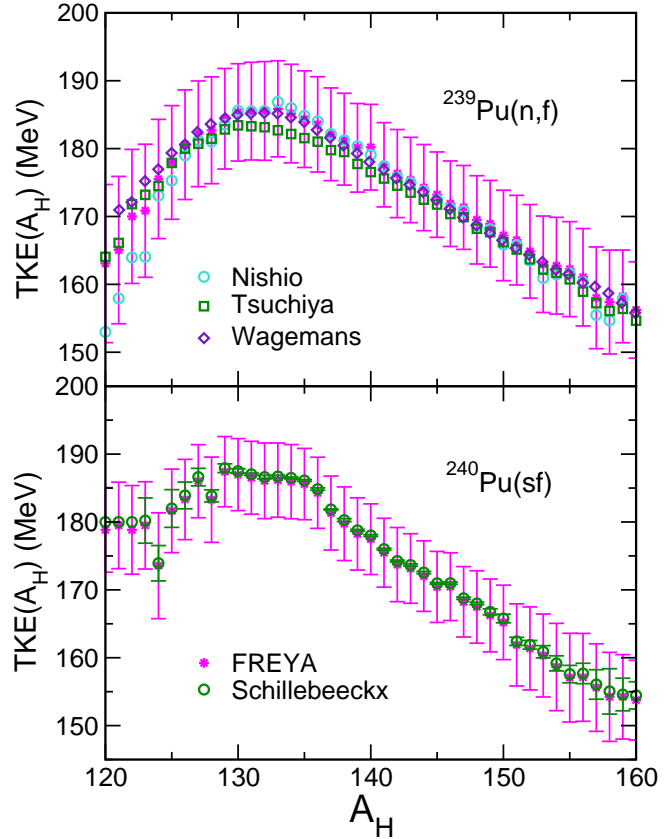}
\caption[]{(Color online) The total fragment kinetic energies as a function 
of the heavy fragment mass for $^{239}$Pu(n$_{\rm th}$,f) 
\protect\cite{NishioPu,TsuchiyaPu,WagemansPu} 
(top) and $^{240}$Pu(sf) \protect\cite{SchillebeeckxPusf} (bottom).  
The \code\ results are shown with the calculated
variance arising from the range of charges available for each $A_H$.}
\label{TKE_Pu_fig}
\end{figure}

\begin{figure}[tbp]
\includegraphics[angle=0,width=\columnwidth]{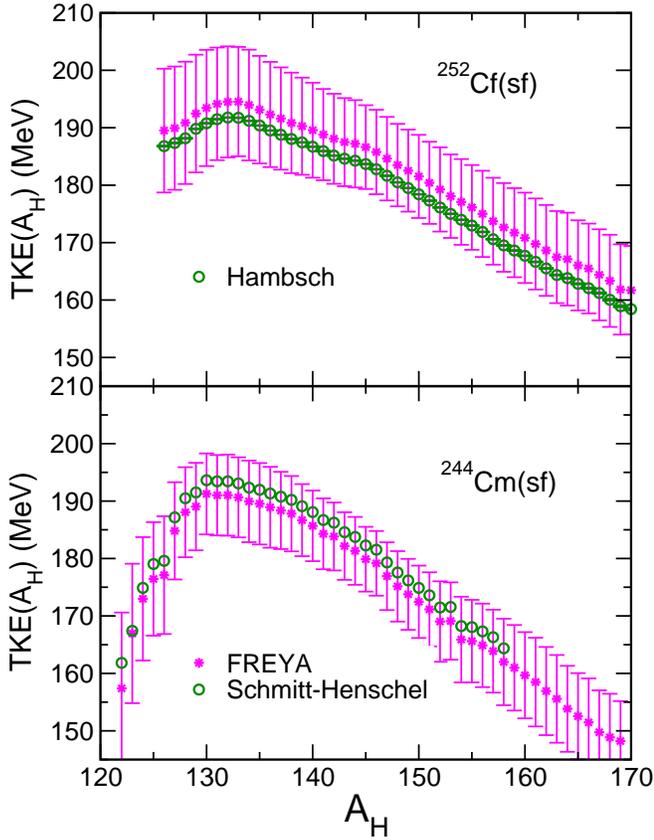}
\caption[]{(Color online) The total fragment kinetic energies as a function 
of the heavy fragment mass for $^{252}$Cf(sf) \protect\cite{HambschCf} (top) 
and $^{244}$Cm(sf) \protect\cite{SchmidtCmTKE} (bottom).  
The \code\ results are shown with the calculated
variance arising from the range of charges available for each $A_H$.}
\label{TKE_CfCm_fig}
\end{figure}

\begin{figure}[tbp]
\includegraphics[angle=0,width=\columnwidth]{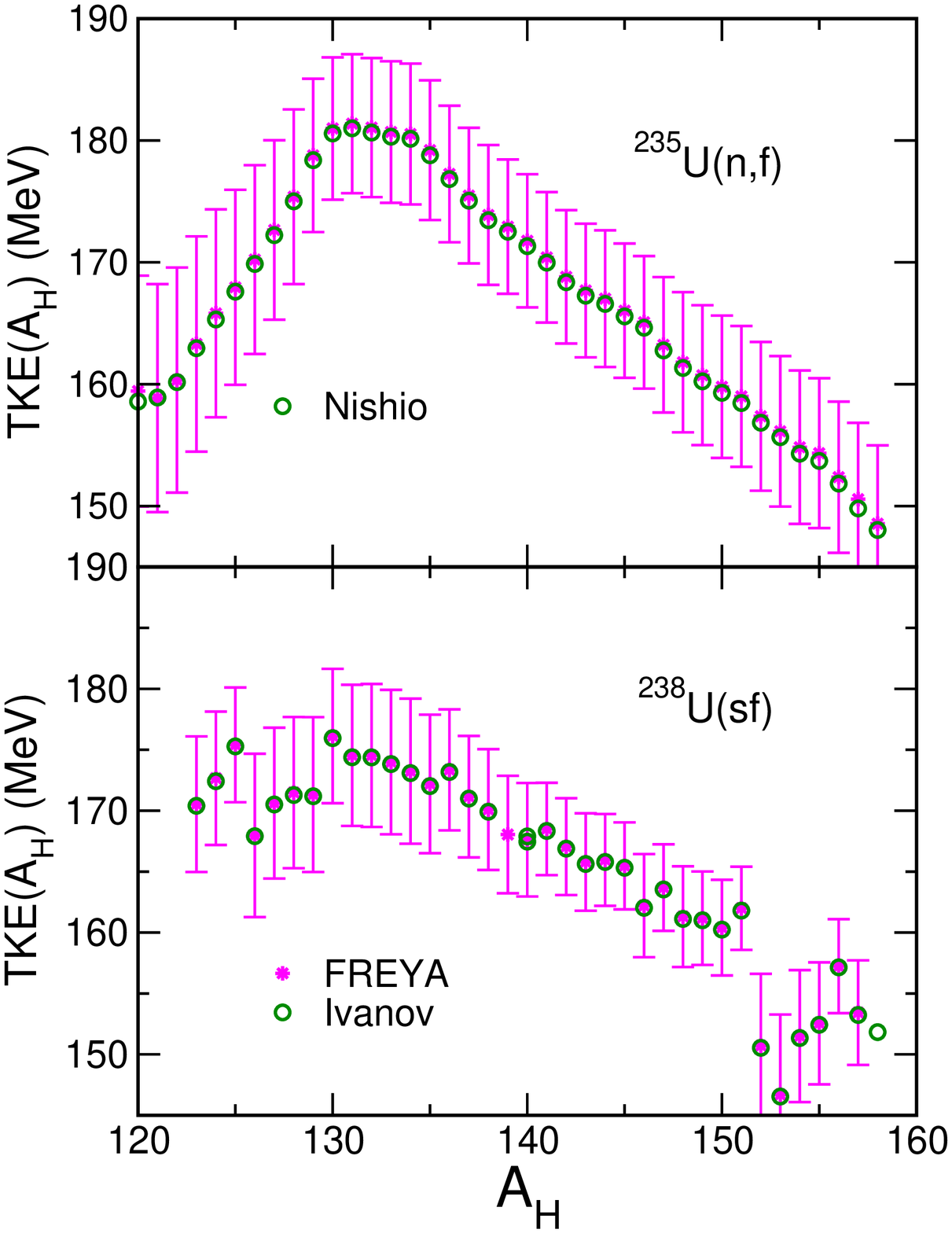}
\caption[]{(Color online) The total fragment kinetic energies as a function 
of the heavy fragment mass for $^{235}$U(n$_{\rm th}$,f) 
\protect\cite{NishioU} (top) and 
$^{238}$U(sf) \protect\cite{IvanovUsf} (bottom).  
The \code\ results are shown with the calculated
variance arising from the range of charges available for each $A_H$.}
\label{TKE_U_fig}
\end{figure}

Figures~\ref{TKE_Pu_fig}-\ref{TKE_U_fig} show the measured average TKE value
as a function of the mass number of the heavy fragment, $A_H$.
Near symmetry, the plutonium fission fragments are mid-shell nuclei
subject to strong deformations.  Thus the scission configuration will contain
significant deformation energy and the TKE will be correspondingly low.  
At $A_H = 132$, the heavy fragment is close to the doubly-magic closed shell 
having $Z_H = 50$ and $N_H = 82$ 
and is therefore resistant to distortions away from sphericity.
Consequently, the scission configuration is fairly compact,
causing the TKE to exhibit a maximum
even though the complementary light fragment is far from a closed shell 
and hence significantly deformed.  Note that the peak around $A_H = 132$ is
a feature of all the data sets shown, regardless of whether fission is neutron
induced or spontaneous and independent of the identity of the fissile nucleus.

The \Punf\ data sets in the top panel of Fig.~\ref{TKE_Pu_fig} 
are very consistent
for $A_H > 135$, above the closed shell at $A_H  = 132$.  In this region and 
below, the agreement among the data sets is not as good, particularly near the
symmetry value of $A_H = 120$, presumably due to the low fragment yields in 
this region.  Unfortunately, no uncertainties are given on the data, only the
full-width half maximum spread of TKE for several given values of $A_H$ in
the measurment of Nishio {\it et al.} \cite{NishioPu}.  This variance is 
similar to that shown for \code.  The \Pusf\ 
data by Schillebeeckx {\it et al.} 
\cite{SchillebeeckxPusf} are somewhat flatter in the region of the closed
shell.  
Unfortunately there are considerable fluctuations in the data for $A_H<130$ 
and TKE($A_H$)was not measured for $A_H < 122$.
We have therefore extrapolated a constant average value back to $A_H = 120$.

The \Cfsf\ data in Fig.~\ref{TKE_CfCm_fig} are again taken from 
Ref.~\cite{HambschCf} with $\cos \theta > 0.9$.  The high statistics of this 
measurement result in small experimental uncertainties and smooth behavior of
TKE$(A_H)$.  There is more uncertainty in the lower statistics \Cmsf\ data
from Ref.~\cite{SchmidtCmTKE}.  Indeed, a comparison of earlier measurements
in that work showed that although the average fragment masses were consistent,
the average TKE of \Cmsf\ varied by 4\% among measurements, depending on
the measurement techniques as well as the choice of calibrators 
(either \Cfsf\ or
$^{235}$U(n$_{\rm th}$,f)).  The results shown in the bottom half of
Fig.~\ref{TKE_CfCm_fig} agree with the highest reported energy and indeed are
$\sim 2$~MeV higher than those reported from \Cfsf\ \cite{SchmidtCmTKE}.  
No more recent results on \Cmsf\ are available.

There are significant fluctuations in the $^{238}$U(sf) data  
\cite{IvanovUsf} in Fig.~\ref{TKE_U_fig}
which can be attributed to the rather low statistics of this
measurement.  We note that the data in Ref.~\cite{IvanovUsf} were presented
as a function of the light fragment mass instead of the heavy.  
The distribution
shown here is obtained by reflection.  We note also that the symmetry region 
is rather poorly measured with points missing around $A_H \approx 119$.  
The  remaining data are at values of TKE below the lower limit of the plot.
Perhaps some of the fluctuations in TKE can be attributed to the `fine 
structure'
noted in the yields in Ref.~\cite{IvanovUsf}.  They also noted that their
measured average TKE was on average 3 MeV lower than their calibrated
result for thermal neutron-induced fission of \Unfnonf.

We assume the average TKE values take the form
\begin{eqnarray}
\overline{\rm TKE}(A_H,E_{\rm n})\ =\ 
	\overline{\rm TKE}_{\rm data}(A_H) + d{\rm TKE}(E_{\rm n})\ .
\label{dtkevalue}
\end{eqnarray}
The first term on the right-hand side of Eq.~(\ref{dtkevalue})
is extracted from the data shown in 
Figs.~\ref{TKE_Pu_fig}-\ref{TKE_U_fig},
while the second term is a parameter adjusted to ensure reproduction
of the measured average neutron multiplicity,
$\overline{\nu}$.
In each particular event, the actual TKE value is then obtained
by adding a thermal fluctuation to the above average, as explained later.

Figures~\ref{TKE_Pu_fig}-\ref{TKE_U_fig} include the average TKE values
calculated with \code\ at thermal energies for neutron-induced fission and
for spontaneous fission,
together with the associated dispersions.  Thus the 
bars associated with the \code\ calculations are {\em not} sampling errors
but indicate the actual width of the TKE distribution for each $A_H$.


\begin{figure}[tbp]
\includegraphics[angle=0,width=\columnwidth]{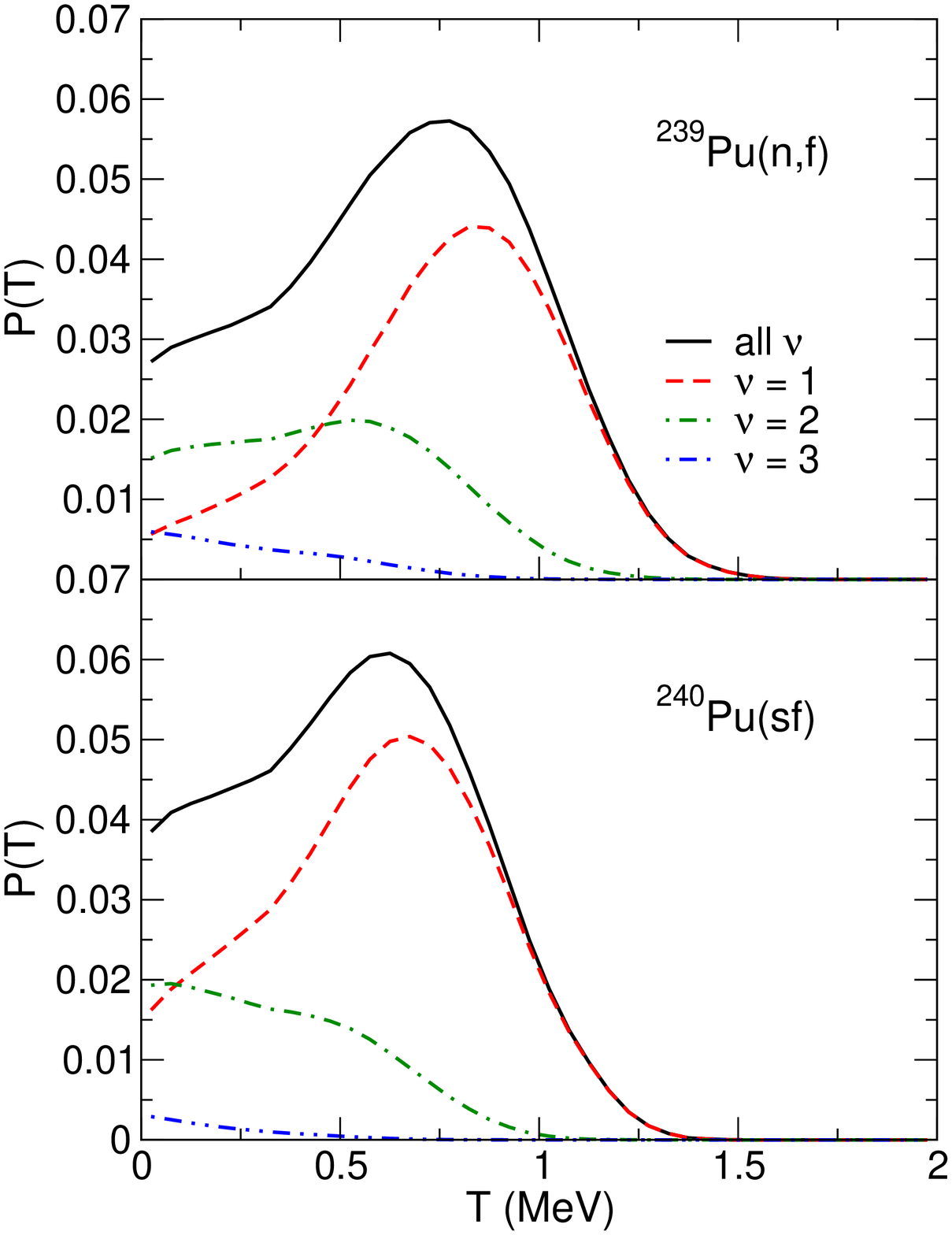}
\caption[]{(Color online) 
The distribution of the maximum temperature in the residual nucleus, $P(T)$,
after $\nu$ neutrons have been evaporated from a primary fragment
for $^{239}$Pu(n$_{\rm th}$,f) (top) and $^{240}$Pu(sf) (bottom).  
The curves show results for $\nu=1$ (dashed), 
$\nu=2$ (dot-dashed), and $\nu=3$ (dot-dot-dashed)
as well as the distribution including all $\nu$ (solid).}
\label{poft_Pu_fig}
\end{figure}

\begin{figure}[tbp]
\includegraphics[angle=0,width=\columnwidth]{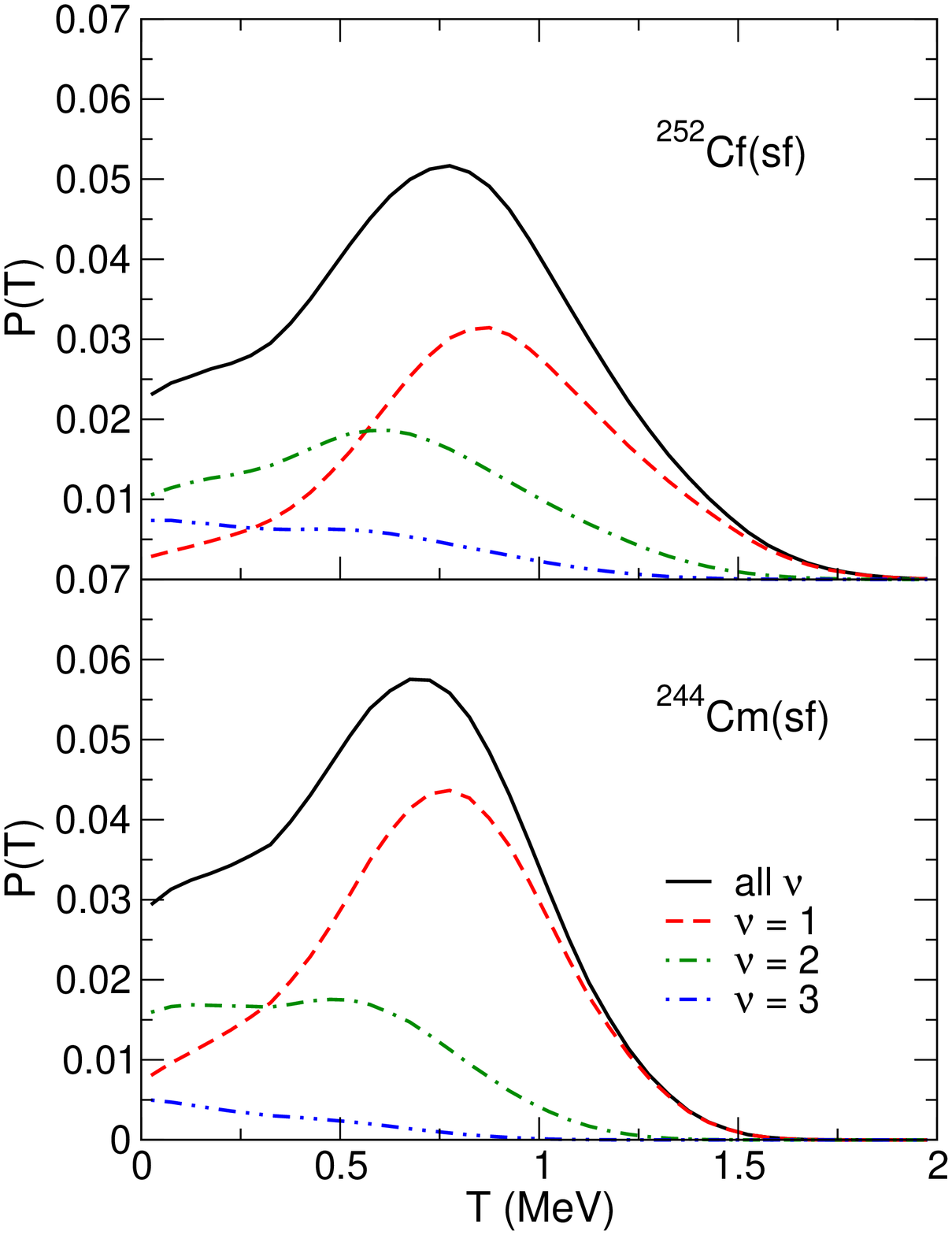}
\caption[]{(Color online) 
The distribution of the maximum temperature in the residual nucleus, $P(T)$,
after $\nu$ neutrons have been evaporated from a primary fragment
for $^{252}$Cf(sf) (top) and $^{244}$Cm(sf) (bottom).  
The curves show results for $\nu=1$ (dashed), 
$\nu=2$ (dot-dashed), and $\nu=3$ (dot-dot-dashed)
as well as the distribution including all $\nu$ (solid).}
\label{poft_CfCm_fig}
\end{figure}

\begin{figure}[tbp]
\includegraphics[angle=0,width=\columnwidth]{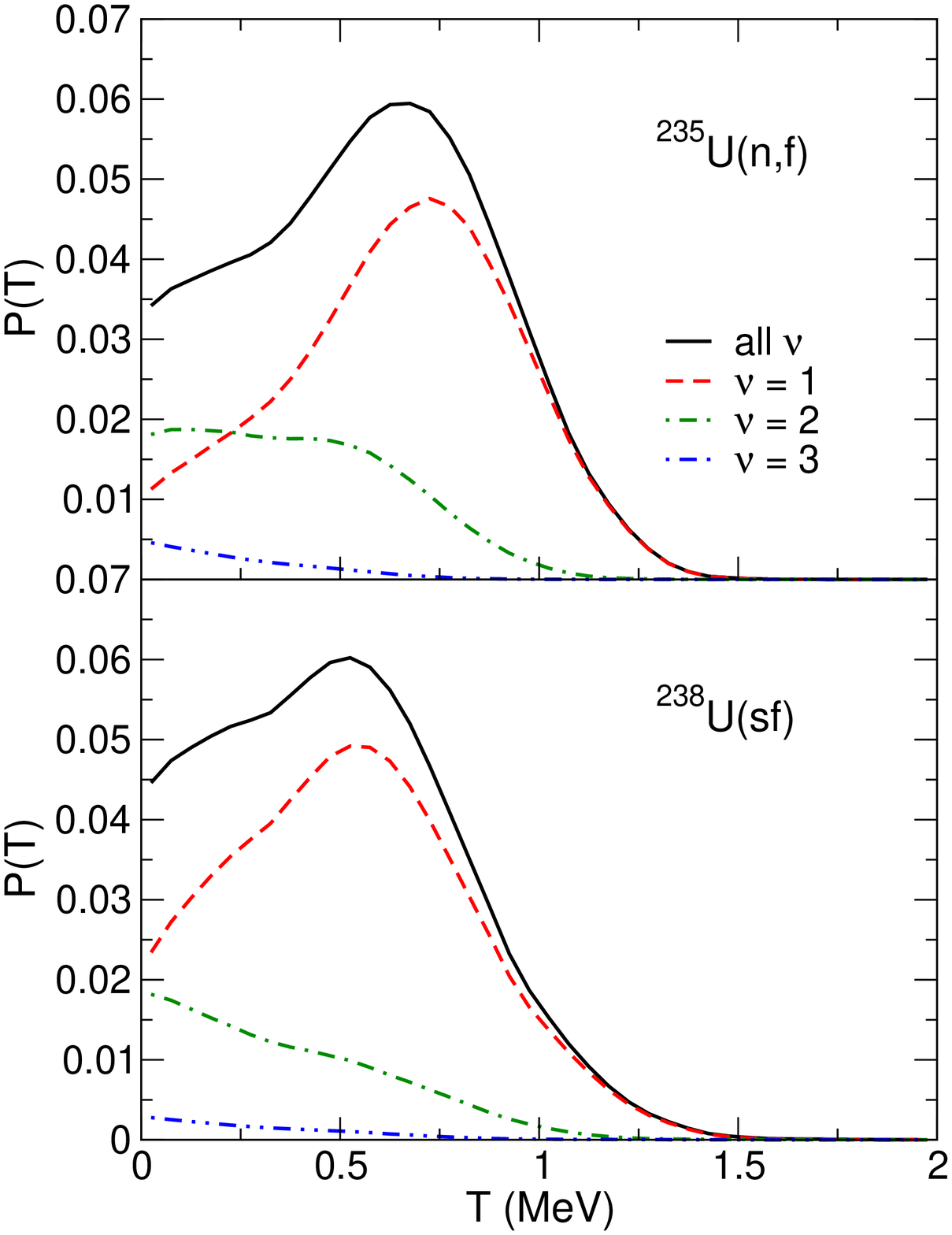}
\caption[]{(Color online) 
The distribution of the maximum temperature in the residual nucleus, $P(T)$,
after $\nu$ neutrons have been evaporated from a primary fragment
for $^{235}$U(n$_{\rm th}$,f) (top) and $^{238}$U(sf) (bottom).  
The curves show results for $\nu=1$ (dashed), 
$\nu=2$ (dot-dashed), and $\nu=3$ (dot-dot-dashed)
as well as the distribution including all $\nu$ (solid).}
\label{poft_U_fig}
\end{figure}

\section{Neutron emission}

Once the average total fragment kinetic energy has been obtained, 
the average combined excitation energy in the two fragments follows
by energy conservation,
\begin{equation}
\overline{\rm TXE}\ =\ \overline{E}_L^*+\overline{E}_H^*\
\doteq\	Q_{LH} - \overline{\rm TKE}\ .
\end{equation}
The first relation indicates that the total excitation energy
is partitioned between the two fragments.
As is common, we assume that the fragment level densities are of the form
$\rho_i(E_i^*)\sim\exp(2\sqrt{{a}_iU_i})$, 
where $U_i$ is the effective statistical energy in the fragment and $a_i$ is
the level-density parameter.  We follow the prescription of Ref.~\cite{VRBDO}
with the value of the asymptotic level density parameter $e_0$ obtained from
the $^{239}$Pu evaluation, assuming it to be universal.

If the two fragments are in mutual thermal equilibrium, $T_L\!=\!T_H$,
the total excitation energy will
be proportional to the level-density parameters,
{\em i.e.}\ $\overline{E}^*_i\sim{a}_i$.
\code\ therefore first assigns tentative average excitations
based on such an equipartition,
\begin{equation}
\acute{E}_i^*\ =\ 
{{a}_i(\tilde{E}_i^*) \over 
{a}_L(\tilde{E}_L^*)+{a}_H(\tilde{E}_H^*)}\,\overline{\rm TXE}\ ,
\end{equation}
where $\tilde{E}_i^*=(A_i/A_0)\overline{\rm TXE}$.
Subsequently, because the observed neutron multiplicities suggest
that the light fragments tends to be disproportionately excited,
the average values are adjusted in favor of the light fragment,
\begin{eqnarray}
\overline{E}^*_L = x \acute{E}^*_L\ ,\ 
\overline{E}^*_H = \overline{\rm TKE}-\overline{E}^*_L\ ,
\label{eeshift}
\end{eqnarray}
where $x$ is an adjustable model parameter expected be larger than unity.

After the mean excitation energies have been assigned,
\code\ considers the effect of thermal fluctuations.
The fragment temperature $T_i$ is obtained from 
$\overline{U}_i\equiv U_i(\bar{E}_i^*)=a_iT_i^2$
and the associated variance in the excitation $E_i^*$ 
is taken as $\sigma_i^2=2\overline{U}_i^*T_i$,
where $U(E^*)=E^*$ in the simple (unshifted) scenario.

Therefore, for each of the two fragments,
we sample a thermal energy fluctuation $\delta E_i^*$ 
from a normal distribution of variance $\sigma_i^2$
and modify the fragment excitations accordingly, so that
\beq
E_i^*\ =\ \overline{E}_i^*+\delta E_i^*\ ,\ i=L, H .
\eeq
Due to energy conservation, there is a compensating opposite fluctuation in
the total kinetic energy \cite{VRBDO}.  The corresponding dispersions 
are included in Figs.~\ref{TKE_Pu_fig}-\ref{TKE_U_fig}.

\subsection{Temperature distributions}

\begin{figure}[tbh]
\includegraphics[angle=0,width=\columnwidth]{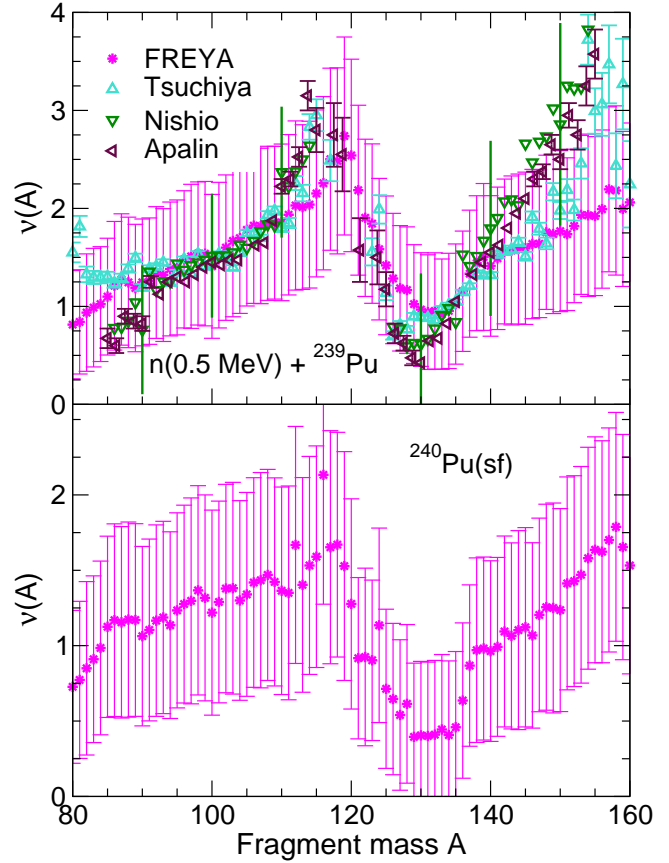}
\caption[]{(Color online) The neutron multiplicity as a function of fragment
mass for neutron-induced fission of $^{239}$Pu 
\protect\cite{NishioPu,TsuchiyaPu,ApalinPu} (top) and 
spontaneous fission of $^{240}$Pu (bottom). 
The vertical bars at intervals of $A=10$ on the data from 
Ref.~\protect\cite{NishioPu} in the top panel
are the full-width half maximum of the 
distribution in neutron multiplicity from the range of charges available for
that $A$.The \code\ results also show this variance by the vertical bars.
(Note that the scales on the $y$-axes of the plots are not identical.)}
\label{nuvsa_pu_fig}
\end{figure}

\code\ assumes that the (fully accelerated) fission fragments
first deexcite by (possibly sequential) neutron evaporation,
followed by sequential photon emission.
At each stage in the decay chain, the spectral shape of the ejectile
is determined by the maximum temperature in the
daughter nucleus, which in turn is determined by the excitation in 
the emitting nucleus and the associated $Q$-value.
Because there are many different $Q$-values involved,
one for each fragment species, 
and becaue the excitation in the emitter fluctuates,
the maximum temperature in the daughter nucleus displays a non-trivial
distribution.

The distributions of the maximum temperature in the evaporation
daughter, granddaughter, and great granddaughter nuclei are shown in
Figs.~\ref{poft_Pu_fig}-\ref{poft_U_fig} for the various cases under study.
The maximum temperature in the daughter nuclei ($\nu$=1)
is fairly well peaked at around 0.5-0.8~{\rm MeV};
the larger contribution comes from the light fragment
which tends to be hotter than its heavy partner.
The temperature distributions in the granddaughter nuclei ($\nu$=2)
are considerably broader and peak at lower energies,
In the case of the great-granddaughters ($\nu$=3) 
the distributions decrease monotonically.
For spontaneous fission of \Pusfnosf\ and \Usfnosf, where the average total 
neutron multiplicity is only $\overline \nu$ is 2.15 and 2.0, respectively, 
it is most likely that each fragment emits just a single neutron
so the probability that a fragment emits three neutrons is rather small,
and the probability for further neutron emission is, in most cases, too 
small to be visible on the plot,
though such events do contribute to the overall distribution near zero.  

The induced fission processes
\Punfnonf(n$_{\rm th}$,f) and \Unfnonf(n$_{\rm th}$,f)
lead to higher daughter temperatures
than the spontaneous processes \Pusfnosf(sf) and \Usfnosf(sf)
and they also yield larger multiplicities.
Thus $\nubar\approx2.88$ for \Punf\ relative to 2.15 for \Pusf,
and $\nubar\approx2.47$ for \Unf\ relative to 2.0 for \Usf.
Relative to \Pusf\ and \Usf,
spontaneous fission of the heavier actinides, $^{244}$Cm and $^{252}$Cf, 
also results in higher daughter temperatures.
Indeed $P(T)$ is similar for \Cmsf\ and \Punf, as are the average neutron
multiplicities, 2.72 and 2.88, respectively. 
We note that for \Cfsf\ the maximum daughter temperature 
peaks at $T \approx 1$~MeV, which is higher than in all the other cases 
considered, and the tail extends up to $T\approx2$~MeV.  
In addition, the distribution of the maximum temperature in the granddaughter
has a distinct peak around $T\approx0.6$~MeV,
which is not surprising considering that $\nubar\approx3.75$ for \Cfsf.

In the treatment by Madland and Nix \cite{MadlandNix} 
a convenient analytical expression was obtained by assuming that the overall 
distribution of the maximum daughter temperature has a triangular shape.
The results in Figs.~\ref{poft_Pu_fig}-\ref{poft_U_fig} demonstrate 
that such an assumption is hardly realistic, even for just $\nu=1$.

\subsection{Average neutron multiplicity}

\begin{figure}[tbp]
\includegraphics[angle=0,width=\columnwidth]{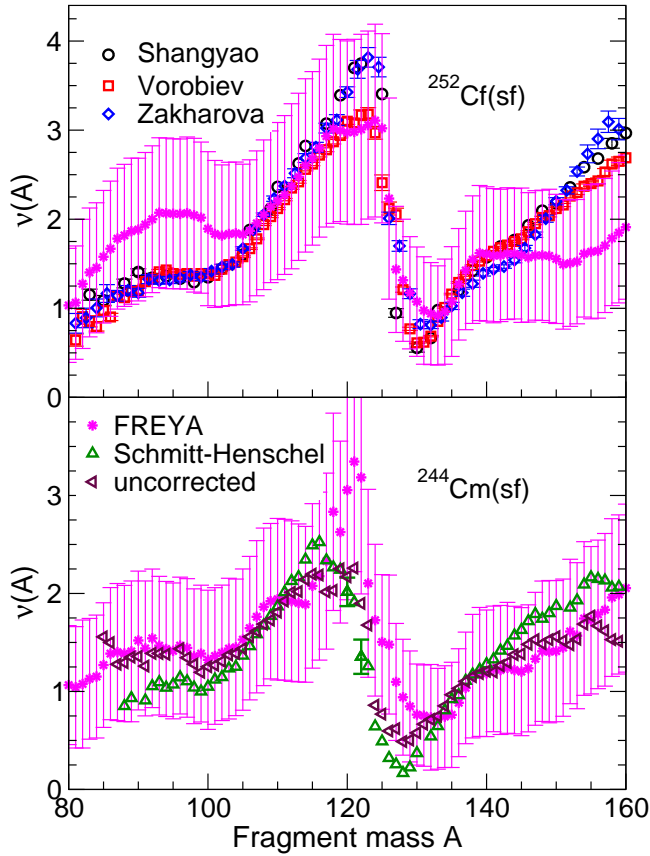}
\caption[]{(Color online) The neutron multiplicity as a function of fragment
mass for spontaneous fission of  $^{252}$Cf 
\protect\cite{ShengyaoCf,VorobievCfnuofa,ZakharovaCf} (top) and $^{244}$Cm 
\protect\cite{SchmidtCmnu} (bottom).  In the bottom part of the plot, the
points labeled uncorrected are also from Ref.~\protect\cite{SchmidtCmnu} but
do not include corrections for fragment motion.  
The \code\ results show the variance in $\nu(A)$ from the range of charges in 
the vertical bars. (Note that the scales on the $y$-axes of the plots are 
not identical.)}
\label{nuvsa_cfcm_fig}
\end{figure}

\begin{figure}[tbp]
\includegraphics[angle=0,width=\columnwidth]{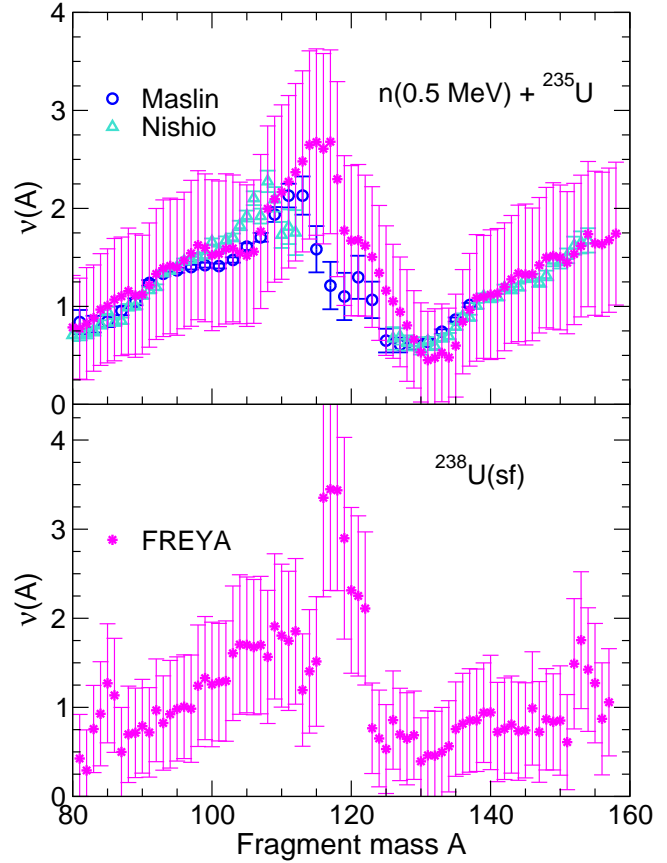}
\caption[]{(Color online) The neutron multiplicity as a function of fragment
mass for neutron-induced fission of $^{235}$U 
\protect\cite{NishioU,MaslinU} (top) and spontaneous fission of $^{238}$U.
The \code\ results show the variance in $\nu(A)$ from the range 
of charges in the vertical bars. (Note that the scales on the $y$-axes of 
the plots are not identical.)}
\label{nuvsa_u_fig}
\end{figure}

The dependence of the average neutron multiplicity
on the fragment mass number $A$, is very sensitive to 
the division of the excitation energy which is governed by the parameter $x$
in Eq.~(\ref{eeshift}). 
As shown in Figs.\ \ref{nuvsa_pu_fig}-\ref{nuvsa_u_fig}, all the measurements
exhibit a characteristic `sawtooth' behavior: the neutron multiplicity from
the light fragment increases slowly as $A$ approaches \half$A_0$ and then drops
rather sharply to a minimum around $A_H \sim 130$, the same location as the
maximum of TKE($A_H$).  Due to the presence of the closed shell at that point,
the fragments are particularly resistent to neutron emission.
Past the dip region, the multiplicity again increases.  The dip tends to be
more sharply defined for larger nuclei where \half$A_0$ is close to 130.  
For example, the drop is particularly abrupt for $^{252}$Cf 
where \half$A_0= 126$.
Where data are available, it is seen that the \code\ calculations provide 
a rather good representation of the `sawtooth' behavior 
of $\nubar(A)$,
even though \code\ is not tuned to these data.  

\begin{figure}[tbp]
\includegraphics[angle=0,width=\columnwidth]{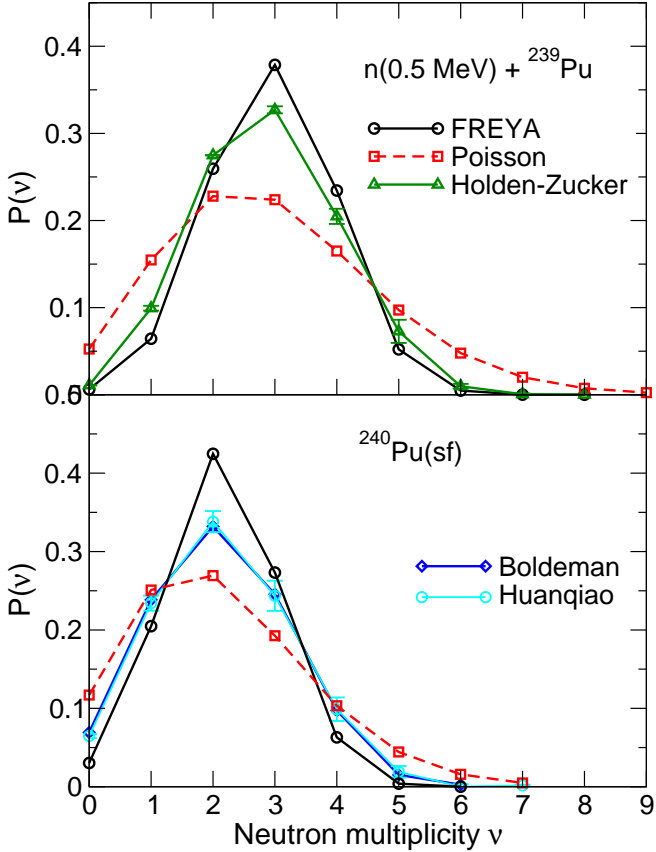}
\caption[]{(Color online) The probability for a given neutron multiplicity 
as a function of multiplicity, $\nu$, for neutron-induced fission of 
$^{239}$Pu \protect\cite{Zuckerpofnu} (top) 
and spontaneous fission of $^{240}$Pu 
\protect\cite{HuanqiaoCmPusf,BoldemanPusf} (bottom). The \code\ results are
shown by the black circles while the equivalent Poisoon distribution is
shown by the red squares.}
\label{pvsnu_Pu_fig}
\end{figure}

\begin{figure}[tbp]
\includegraphics[angle=0,width=\columnwidth]{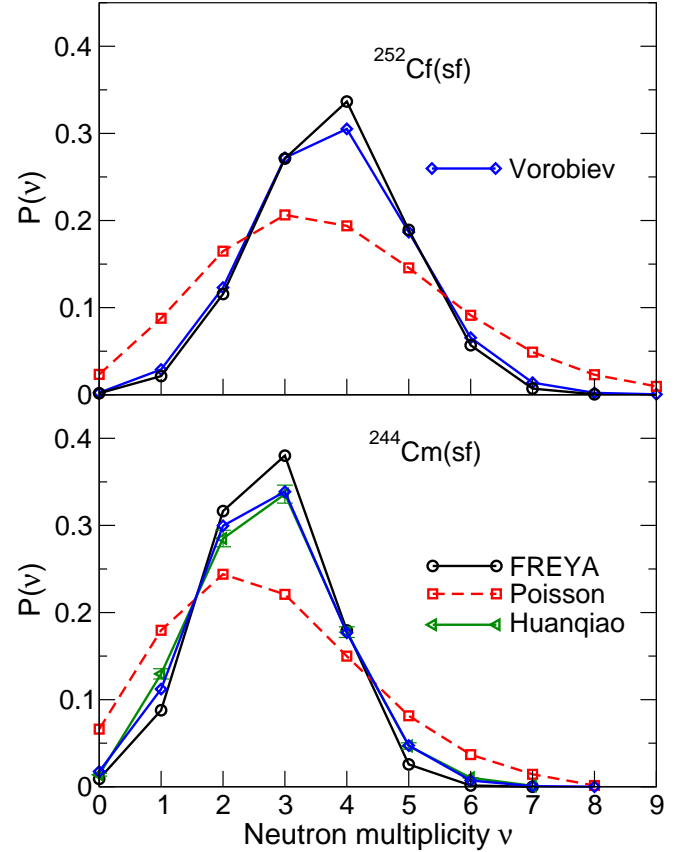}
\caption[]{(Color online) The probability for a given neutron multiplicity 
as a function of multiplicity, $\nu$, for spontaneous fission of 
$^{252}$Cf \protect\cite{VorobievCfpofnu} (top) and 
$^{244}$Cm \protect\cite{HuanqiaoCmPusf,VorobievCmpofnu} (bottom). 
The \code\ results are
shown by the black circles while the equivalent Poisoon distribution is
shown by the red squares.}
\label{pvsnu_CfCm_fig}
\end{figure}

Although the agreement
is generally good, the observed behavior is not perfectly reproduced.
The \code\ results for \Punfnonf(n$_{\rm th}$,f) in 
Fig.~\ref{nuvsa_pu_fig} agree very well with the data
for $90 < A< 140$ with $x = 1.1$.  At higher and lower $A$, although
there are deviations, the measurements are within the variance of the \code\
results.  However, in these regions, as well as near symmetry, the yields
are smaller so that larger deviations may be expected.  
No neutron measurements were made in Ref.~\cite{SchillebeeckxPusf}
and we have not located any comparison data for \Pusf; here we use $x = 1.2$.
Below the symmetry point, 
the slope of $\nubar(A)$ is rather small, but it increases more rapidly
above $A = 132$.  
The \code\ results display larger fluctuations for this isotope,
which may be attributed to the more irregular behavior of TKE$(A_H)$.  
Also note that while the central values of $\nu(A)$ fluctuate 
more, the variances are the same size as for \Punf\ in the top panel and 
appear exaggerated
by the smaller scale of $\nubar(A)$  on the $y$-axis for \Pusf.

There are numerous measurements of $\nubar(A)$ for \Cfsf.  A sample of some 
representative, more recent, results are shown in the top part of 
Fig.~\ref{nuvsa_cfcm_fig}.  The measurements are all very similar with
small differences only near $A< 90$, $A \sim 120$, and $A>150$.  The light
fragment data are rather flat and then increase rather quickly for $105 < A <
120$.  Above $A = 132$, the slope of $\nubar(A)$ is less than that seen for
\Punf\ in Fig.~\ref{nuvsa_pu_fig}.  The \code\ results, obtained with 
$x = 1.3$, while consistent within the variance of the data, are 
significantly flatter than the data for $A < 100$ and $A > 140$. 
This behavior, stronger than any of the other \code\ results in 
Figs.~\ref{nuvsa_pu_fig}-\ref{nuvsa_u_fig}, can be traced to the apparent
two-slope behavior of TKE($A_H$) for \Cfsf\ in Fig.~\ref{TKE_CfCm_fig}: there
is a slow decrease in TKE for $132 < A_H < 145$ with a faster change of TKE
with $A_H$ thereafter.  It is also consistent with the large widths of the
asymmetric fission yields in Fig.~\ref{yields_CfCm_fig}.  We note that changing
$x$ does not change the slopes of $\nubar(A)$, only the relative magnitudes.
To better describe $\nubar(A)$ with \code, it would be necessary to fit
$x(A)$ rather than employing just a single-valued parameter for $x$.

Data for \Cmsf\ are shown in the lower panel of Fig.~\ref{nuvsa_cfcm_fig}.
Both data sets shown are from the same experiment \cite{SchmidtCmnu}.  
Two results were given
in the paper due to the location of the fission source relative to the neutron
detector, the corrected version, labeled Schmidt-Henschel, shifts the measured
$\nubar(A)$ to account for the fact that 
only some of the neutrons emitted by the fragments will reach the detector.  
The authors calibrated their correction for \Cfsf\ by normalizing the position 
of the sawtooth with $A$ to previously published data.  
The \Cmsf\ correction was made by scaling the \Cfsfnosf\ shift by the
ratio of the total neutron multiplicity in \Cmsfnosf\ relative to \Cfsfnosf,
presumably because they could not compare their \Cmsfnosf\ results to
other measurements of the same system.  The authors were rather inconclusive
about which results were actually correct since the sum of complementary 
multiplicities, $\nubar(A) + \nubar(A_0 - A)$, did not agree well
with the total neutron multiplcity, $\nubar_T(A)$ near \half$A_0$.  If this
disagreement is real, then the correction had the effect of shifting the peak
of $\nubar(A)$ downward from $A\sim 120$ to $A \sim 117$. 
Interestingly, the \code\ results, calculated with $x = 1.2$ and
treating the motion of both the fragments and the neutrons relativistically, 
agrees much better with the uncorrected results of Ref.~\cite{SchmidtCmnu}.

Finally, \code\ results are compared to \Unf\ in the top panel of 
Fig.~\ref{nuvsa_u_fig}.  The values of $\nubar(A)$ agree very well with the
sawtooth pattern of the data \cite{NishioU,MaslinU} with the exception of
the symmetric region where the yields are rather low.  
Indeed,  Ref.~\cite{NishioU} does not provide results for $\nubar(A)$ 
in the symmetric  region.  
The \code\ results for \Usf, calculated with $x = 1.2$, 
are shown in the bottom panel of the figure.  
The shape of the sawtooth appears rather flat for this isotope,
likely because of the high $\nubar(A)$ obtained near symmetry, reflecting the
low TKE reported by Ref.~\cite{IvanovUsf} in Fig.~\ref{TKE_U_fig}.

\subsection{Neutron multiplicity distribution}

\begin{figure}[tbp]
\includegraphics[angle=0,width=\columnwidth]{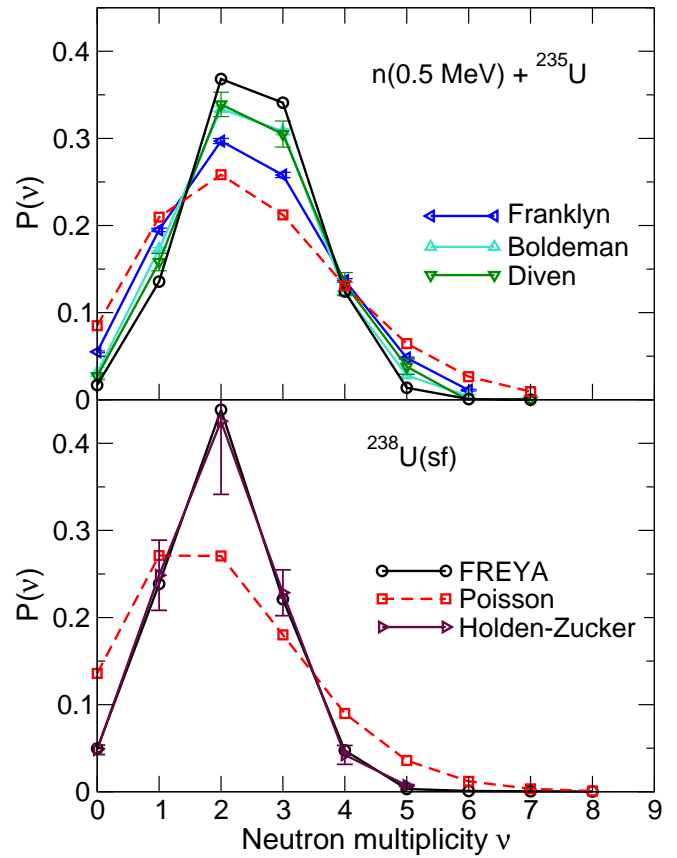}
\caption[]{(Color online) 
The probability for a given neutron multiplicity 
as a function of multiplicity, $\nu$, for neutron-induced fission of 
$^{235}$U \protect\cite{FranklynU,BoldemanU,Diven} (top) and spontaneous 
fission of $^{238}$U  \protect\cite{Zuckerpofnu} (bottom). 
The \code\ results are
shown by the black circles while the equivalent Poisoon distribution is
shown by the red squares. }
\label{pvsnu_U_fig}
\end{figure}

\begin{figure}[tbp]
\includegraphics[angle=0,width=\columnwidth]{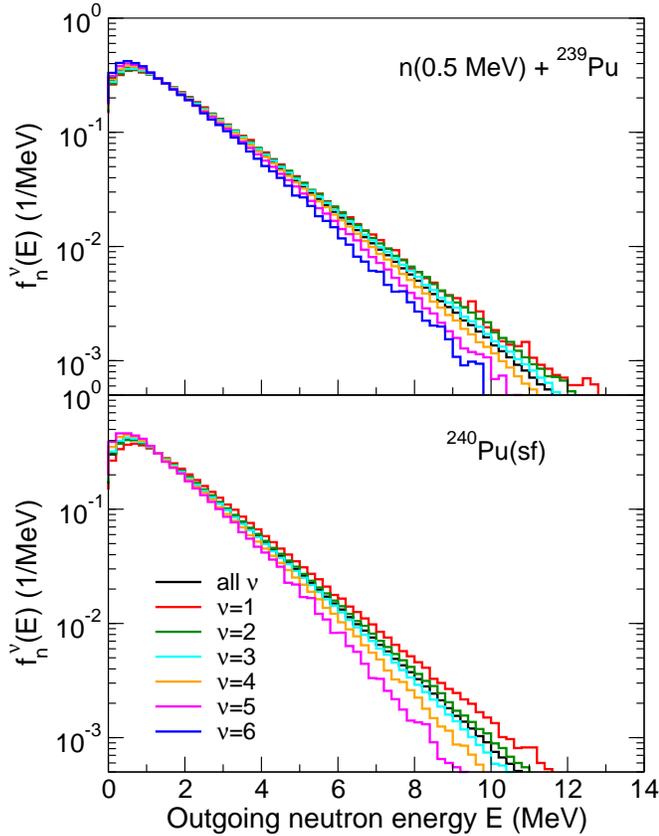}
\caption[]{(Color online) 
The spectral shapes of prompt fission neutrons
for neutron-induced fission of $^{239}$Pu (top) and 
spontaneous fission of $^{240}$Pu (bottom),
for events with a fixed neutron multiplicity of
$\nu=1,\dots,6$ as well as for all events
irrespective of the associated multiplicity.}
\label{spectra_Pu_fig}
\end{figure}

Figures~\ref{pvsnu_Pu_fig}-\ref{pvsnu_U_fig} show 
the neutron multiplicity distribution $P(\nu)$
for the various cases considered.
Each emitted neutron reduces the excitation energy in the residue 
by not only its kinetic energy (recall $\overline E = 2T$ 
where $T$ is the maximum temperature in the daughter nucleus)
but also by the separation energy $S_{\rm n}$ 
(which is generally significantly larger).
Therefore the resulting $P(\nu)$ is narrower than
a Poisson distribution with the same average multiplicity, 
as clearly seen in the figures.  

In experiments, the quantity $P(\nu)$ is determined by detecting fission 
events in a sample of material and correlating these with simultaneous 
neutron detection.
The relative probability for emission of $\nu$ neutrons in  given event,
$P(\nu)$, is inferred by combining the calculated probability 
for observing $n$ neutrons when $\nu$ were emitted, $Q(n;\nu)$,
with the detector efficiency determined from the count rate 
by comparison with a calibration source having a known $\nubar$; 
typically \Cfsf\ is used. 
Thus, while the value of $\nubar$ may be well measured for a given isotope, 
the distribution $P(\nu)$ is less well determined.

We compare to data in so far as possible.  The results labeled `Holden-Zucker'
in Figs.~\ref{pvsnu_Pu_fig} for \Punf\ and \ref{pvsnu_U_fig} for \Usf\ are
consensus values from a 1985 report by Holden and Zucker \cite{Zuckerpofnu}.
Results from this reference are generally available for the other cases
shown here.  However, we do not show them if there is good agreement between
the data displayed here and in Ref.~\cite{Zuckerpofnu} 
or if there are more recent
data available.  In most cases, the agreement is rather good, 
with the \code\ results following the data more closely 
than the equivalent Poisson distribution.

The largest difference between the calculated \code\ multiplicity distributions
and the data seems to be for \Pusf, 
which may be due to the smaller sample of \Pusf\ $Y(A)$
and TKE($A_H$) data used as input to \code.  
By contrast, the agreement of \code\ with the \Usf\ consensus data 
given in Ref.~\cite{Zuckerpofnu} is remarkable.

\subsection{Multiplicity-gated neutron spectra}

\begin{figure}[tbp]
\includegraphics[angle=0,width=\columnwidth]{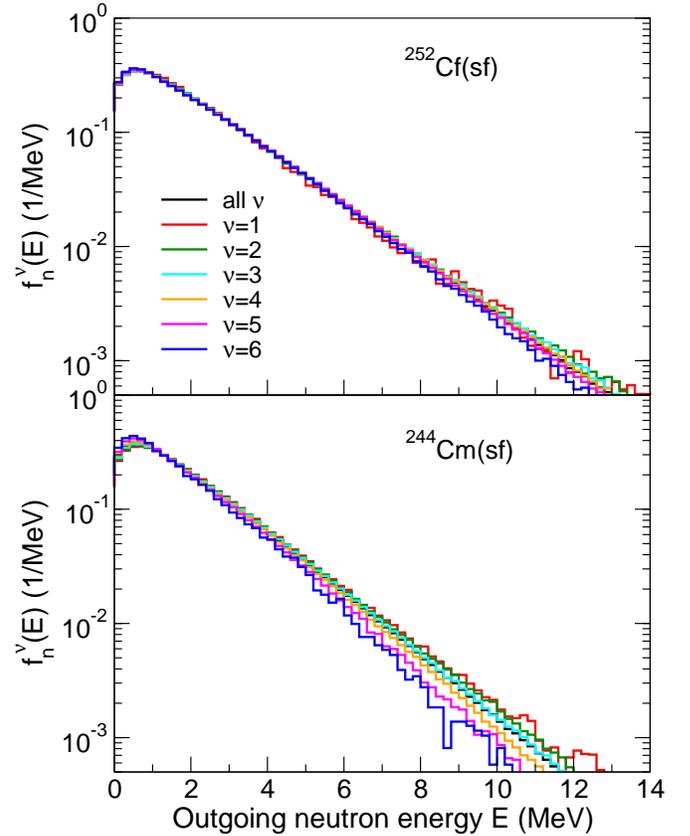}
\caption[]{(Color online) 
The spectral shapes of prompt fission neutrons
for spontaneous fission of  $^{252}$Cf (top) and $^{244}$Cm (bottom),
for events with a fixed neutron multiplicity of
$\nu=1,\dots,6$ as well as for all events
irrespective of the associated multiplicity.}
\label{spectra_CfCm_fig}
\end{figure}

\begin{figure}[tbp]
\includegraphics[angle=0,width=\columnwidth]{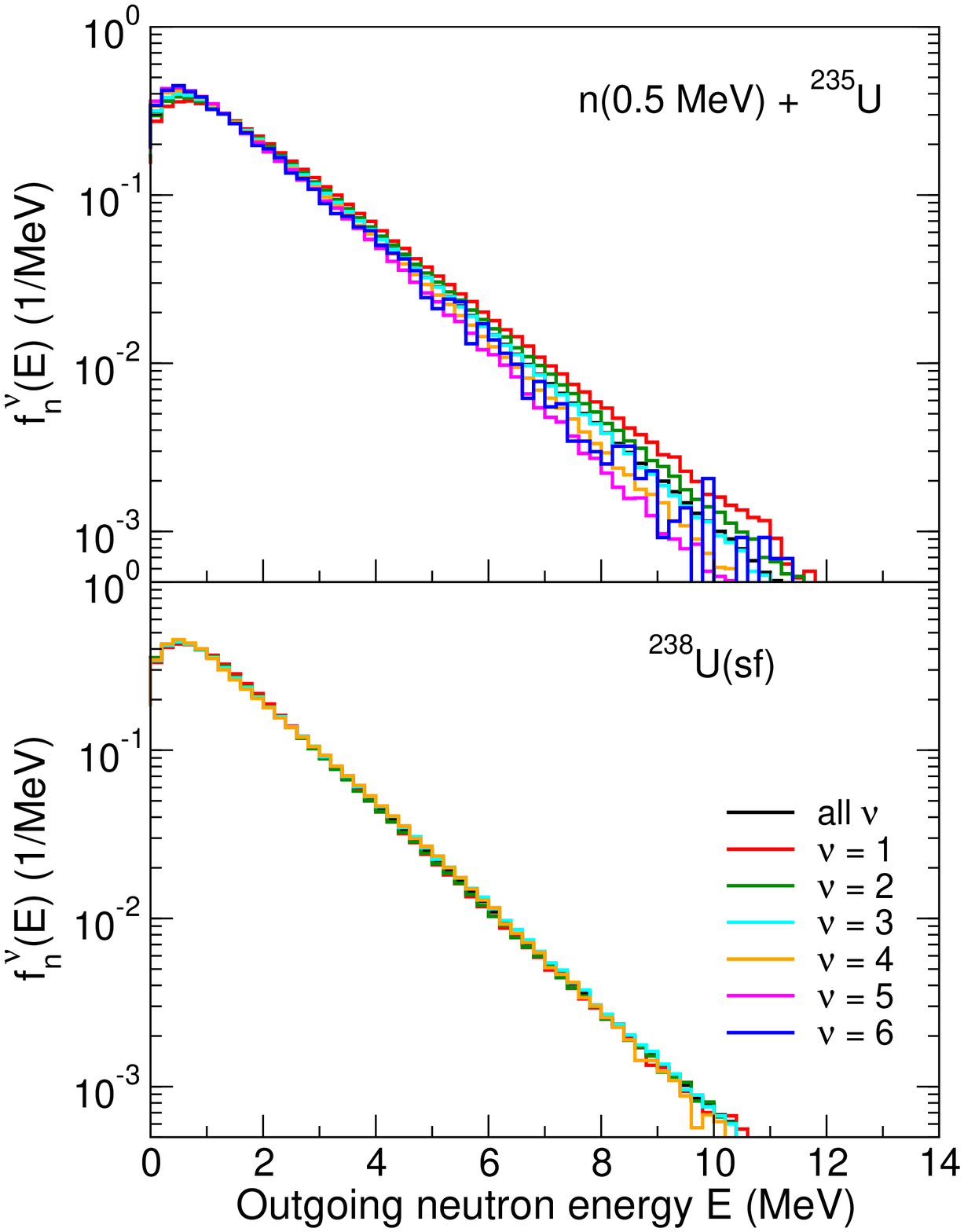}
\caption[]{(Color online) 
The spectral shapes of prompt fission neutrons
for neutron-induced fission of $^{235}$U (top) and 
spontaneous fission of $^{238}$U (bottom),
for events with a fixed neutron multiplicity of
$\nu=1,\dots,6$ as well as for all events
irrespective of the associated multiplicity.}
\label{spectra_U_fig}
\end{figure}

For fisison events having a specified total neutron multiplicity $\nu$,
we define the associated spectral {\em shape},
\begin{equation}
f_{\rm n}^\nu(E)\ \equiv\ {1\over\nu}{d\nu\over dE}\ ,
\end{equation}
which is thus normalized to unity, while
the corresponding spectral shape of the neutrons from all the fission events
irrespective of the associated multiplicity is denoted simply by $f_{\rm n}(E)$
and is also normalized to unity.

The multiplicity-gated spectral shapes obtained for the various
cases considered are shown in Figs.\ \ref{spectra_Pu_fig}-\ref{spectra_U_fig}.
Results are presented for multiplicities up to $\nu = 6$.
It is apparent that the spectra become progressively softer 
at higher multiplicities, as one would expect because more neutrons
are sharing the available energy.
This type of elementary conservation-based correlation feature
is not provided by the standard models of fission.

The tails of the prompt fission neutron spectra from \Pusf\ are longer and
broader than
those from \Punf\ even though the average energies are smaller and fewer
neutrons are emitted.  The opposite is the case for \Usf\ and \Unf\ since
the spectra from \Usf\ are closely clustered around the mean.  The most 
energetic neutrons at high multiplicity are emitted from \Cfsf\ where the
spectra are also rather closely clustered around the mean. The spectral
shapes and average energies of \Cmsf are rather similar to \Punf.
  
Table \ref{aveEvsn} shows the mean
kinetic energy of neutrons emitted from 
the two fragments
as a function of the neutron multiplicity.  The averages are shown for
multiplicities with event samples large enough to prevent the results from
being skewed by fluctuations.  The largest average energies are seen for
\Cfsf\ with similar results for \Punf\ and \Cmsf, not surprising given the 
temperature profiles seen in
Figs.~\ref{poft_Pu_fig}-\ref{poft_U_fig}.  The variances are on the same order
as the averages, albeit somewhat smaller, in most cases.

\begin{table}
\begin{center}
\begin{tabular}{|c|cc|cc|} \hline
$\nu$ & $\langle E \rangle$ (MeV) & $\sigma_E$ (MeV)  & $\langle E \rangle$ 
(MeV) & $\sigma_E$ (MeV)  \\ \hline
& \multicolumn{2}{|c|}{n$(0.5\,{\rm MeV})+ {\rm ^{239}Pu}$} & 
\multicolumn{2}{|c|}{$^{240}$Pu(sf)}
\\   \hline
all & 2.06 & 1.68 & 1.86 & 1.52 \\
1   & 2.16 & 1.78 & 2.03 & 1.64 \\
2   & 2.14 & 1.76 & 1.90 & 1.56 \\
3   & 2.09 & 1.71 & 1.83 & 1.49 \\
4   & 2.01 & 1.64 & 1.73 & 1.41 \\
5   & 1.92 & 1.56 & 1.64 & 1.31 \\
6   & 1.84 & 1.48 & 1.52 & 1.19 \\ \hline \hline
& \multicolumn{2}{|c|}{$^{252}$Cf(sf)} & \multicolumn{2}{|c|}{$^{244}$Cm(sf)}
\\   \hline
all & 2.23 & 1.88 & 2.03 & 1.68 \\
1   & 2.21 & 1.87 & 2.13 & 1.77 \\
2   & 2.24 & 1.91 & 2.10 & 1.73 \\
3   & 2.25 & 1.90 & 2.05 & 1.69 \\
4   & 2.24 & 1.88 & 1.98 & 1.62 \\
5   & 2.21 & 1.86 & 1.89 & 1.55 \\
6   & 2.18 & 1.82 & 1.81 & 1.50 \\
7   & 2.14 & 1.78 &  -   &  -   \\
8   & 2.06 & 1.70 &  -   &  -   \\
\hline \hline
& \multicolumn{2}{|c|}{n$(0.5\,{\rm MeV})+ {\rm ^{235} U}$} & 
\multicolumn{2}{|c|}{$^{238}$U(sf)}
\\   \hline
all & 1.92 & 1.59 & 1.76 & 1.46 \\
1   & 2.08 & 1.71 & 1.76 & 1.44 \\
2   & 1.98 & 1.58 & 1.74 & 1.45 \\
3   & 1.92 & 1.50 & 1.77 & 1.47 \\
4   & 1.83 & 1.45 & 1.76 & 1.46 \\
5   & 1.76 & 1.46 &  -   &  -   \\
\hline
\end{tabular} 
\end{center}
\caption[]{The mean neutron kinetic energy, $\langle E \rangle$, 
together with the associated dispersion, $\sigma_E$, 
for events with a fixed neutron multiplicity $\nu$ as well as for all events.}
\label{aveEvsn}
\end{table}

\subsection{Neutron-neutron angular correlations}

The event-by-event nature of \code\ makes it straightforward to
extract the angular correlation between two evaporated neutrons,
an observable that has long been of experimental interest
(see, for example, Refs.\ \cite{DeBenedettiPR74,FranklynU,GagarskiBRAS72}
and references therein)
but which cannot be addressed with the standard models of fission.

\begin{figure}[tbp]
\includegraphics[angle=0,width=\columnwidth]{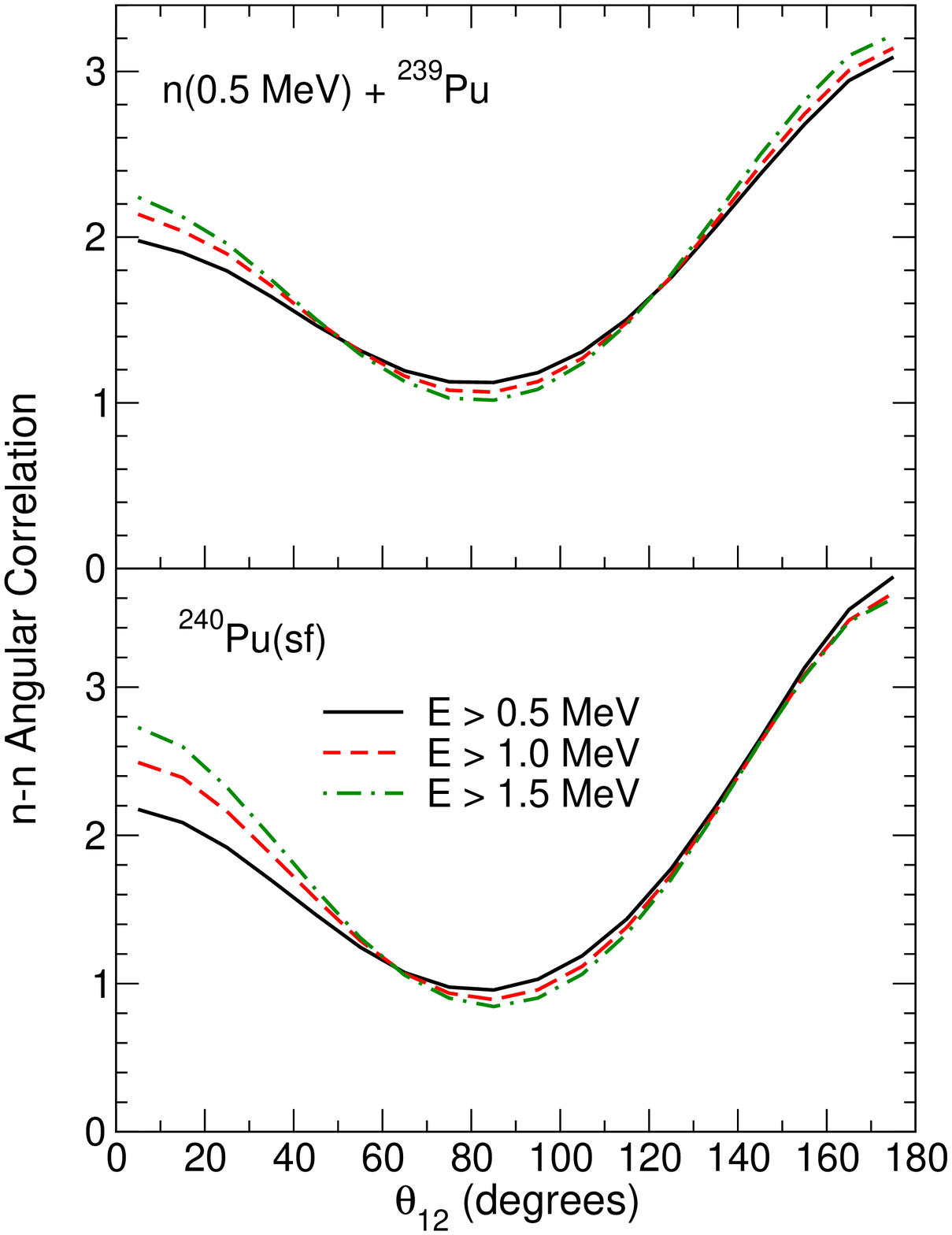}
\caption[]{(Color online) The angular correlation between two neutrons
emitted from $^{239}$Pu(n$_{\rm n}$,f) (top) and $^{240}$Pu(sf) (bottom) 
as a function of the opening angle between the two
neutrons, $\theta_{12}$.  The \code\ results are shown for several cuts on the
neutron kinetic energy: $E>0.5$~MeV (solid black), 1 MeV (dashed red),
and 1.5 MeV (dot-dashed green).}
\label{nncorr_Pu_fig}
\end{figure}

\begin{figure}[tbp]
\includegraphics[angle=0,width=\columnwidth]{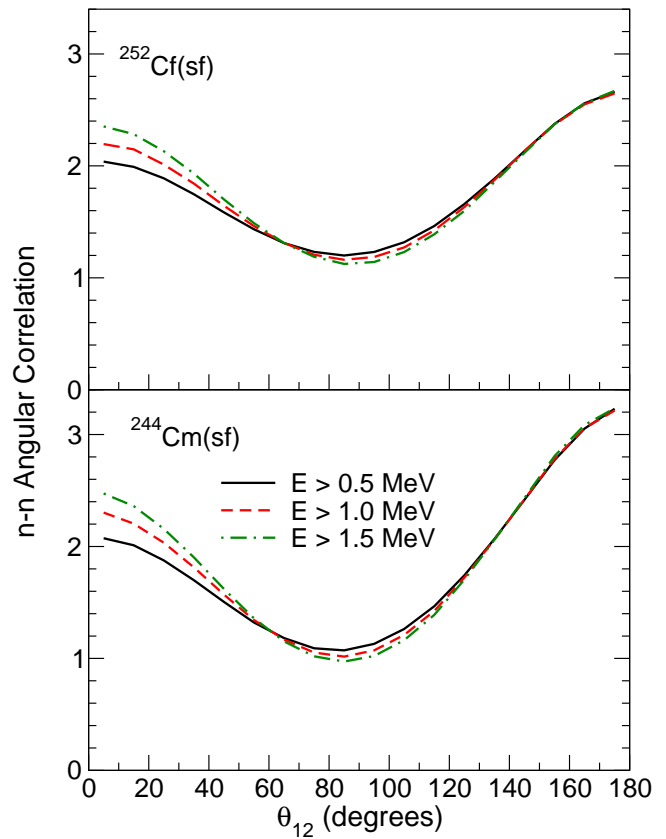}
\caption[]{(Color online) The angular correlation between two neutrons emitted
in spontaneous fission of $^{252}$Cf (top) and
$^{244}$Cm (bottom) as a function of the opening angle between the two
neutrons, $\theta_{12}$.  The \code\ results are shown for several cuts on the
neutron kinetic energy: $E> 0.5$~MeV (solid black), 1 MeV (dashed red),
and 1.5 MeV (dot-dashed green).}
\label{nncorr_CfCm_fig}
\end{figure}

\begin{figure}[tbp]
\includegraphics[angle=0,width=\columnwidth]{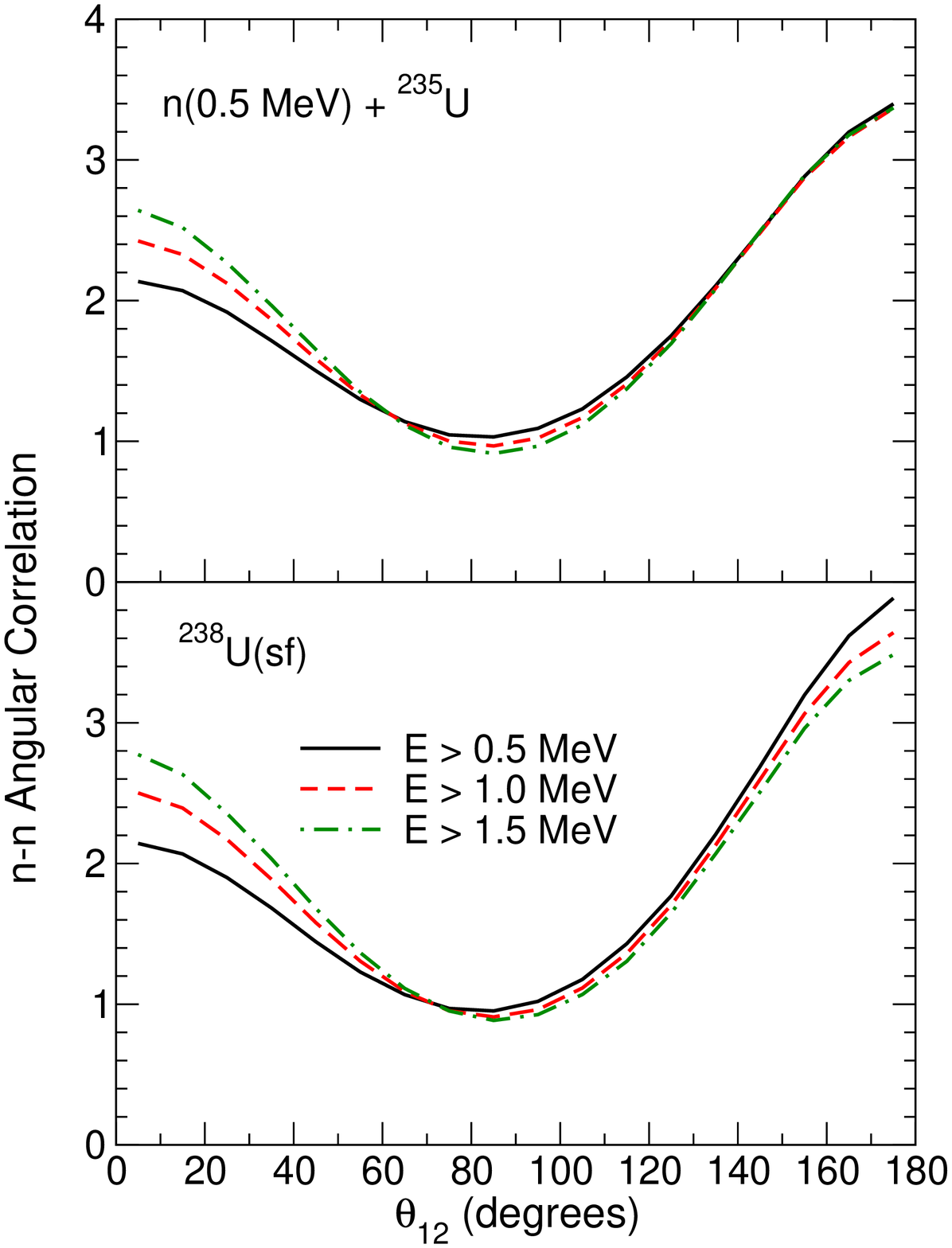}
\caption[]{(Color online) The angular correlation between two neutrons emitted
from $^{235}$U(n$_{\rm n}$,f) (top) and $^{238}$U(sf) (bottom) 
as a function of the opening angle between the two
neutrons, $\theta_{12}$.  The \code\ results are shown for several cuts on the
neutron kinetic energy: 
$E > 0.5$~MeV (solid black), 1 MeV (dashed red),
and 1.5 MeV (dot-dashed green).}
\label{nncorr_U_fig}
\end{figure}

Figures~\ref{nncorr_Pu_fig}-\ref{nncorr_U_fig} show 
this quantity for the neutrons resulting from
fission induced by thermal neutrons on $^{235}$U and $^{239}$Pu as well
as neutron correlations in spontaneous fission.
The results are shown for neutrons with kinetic energies
above thresholds at $E = 0.5$, 1 and 1.5~MeV.
The angular modulation grows somewhat more pronounced as the threshold 
is raised  (while the statistics are correspondingly reduced).

The neutrons tend to be either forward or backward correlated.
The backward correlation appears to be somewhat favored.
While not shown, we have previously 
analyzed the case of \Punf\ for $\nu = 2$, breaking
it down to three separate contributions: both neutrons from the light fragment,
both from the heavy fragment, and one neutron emitted from each fragment
\cite{Sinaiaproc}.  There is a significant correlation at $\theta_{12}=0$ 
when both neutrons are emitted from the same fragment, with a higher peak for
the case when both neutrons are emitted from the light fragment 
due to its higher velocity.  On the other hand, when one
neutron is emitted from each fragment, their direction tends to be
 anti-correlated due to the relative motion of the emitting fragments, 
resulting in a peak at $\theta_{12} = 180$.
The overall result is a stronger backward correlation 
because emission from both fragments is most likely.

The backward correlation is strongest when the overall neutron multiplicity
is low, especially for \Pusf\ and \Usf, whereas large multiplicities,
as for \Cfsf\ and \Cmsf, reduce the angular correlation.

\begin{figure}[b]
\includegraphics[angle=0,width=\columnwidth]{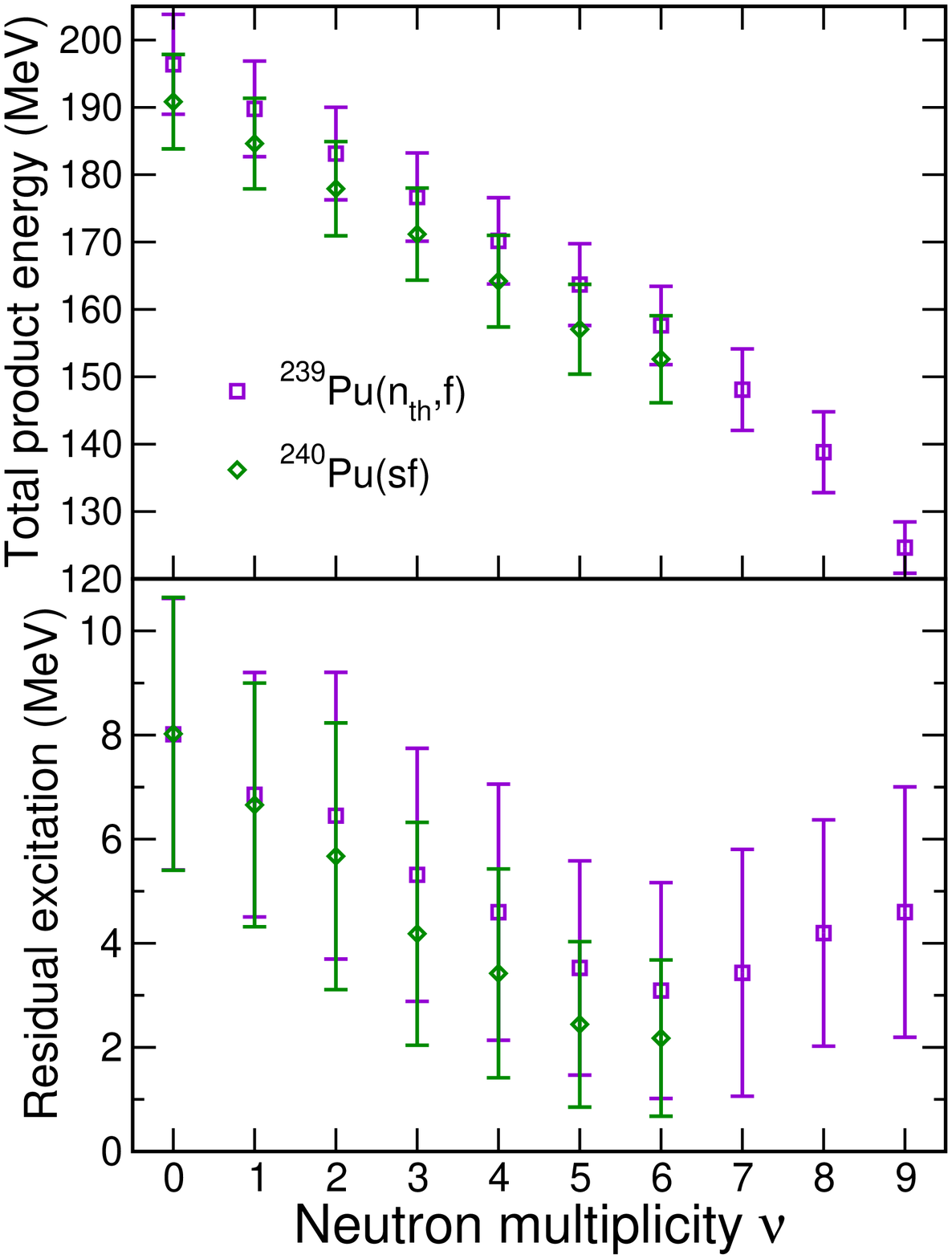}
\caption[]{(Color online) The total product kinetic energy (top) and 
residual excitation energy (bottom) remaining after neutron emission has 
ceased as a function of neutron multiplicity.  The \code\ results are shown
for $^{239}$Pu(n$_{\rm th}$,f) (squares) and $^{240}$Pu(sf) (diamonds).}
\label{prodvsnu_Pu_fig}
\end{figure}

\subsection{Correlations between product energies and neutron multiplicity}

The combined kinetic energy of the two resulting (post-evaporation) 
product nuclei is shown as a function of the neutron multiplicity $\nu$
in the top panels of 
Figs.~22-24. 
It decreases with increasing multiplicity,
as one might expect on the grounds that the emission of more neutrons
tends to carry off more initial excitation energy,
thus leaving less available for the products.
As expected from the behavior of $Z_LZ_H$,
the combined product kinetic energy is largest for the most massive 
fission systems (\Cfsfnosf\ and \Cmsfnosf) 
and lowest for the least massive (\Unfnonf\ and \Usfnosf).

The bottom panels of Figs.~22-24 
show the neutron multiplicity dependence of the
average residual excitation energy in those post-evaporation product nuclei.
Because this energy is available for subsequent photon emission,
one may expect that the resulting photon multiplicity
would display a qualitatively similar behavior and thus,
in particular, be anti-correlated with the neutron multiplicity.

There is little sensitivity of the residual excitation 
to the identity of the fissioning nucleus in any of the cases presented.  
This result shows that the energies left over after prompt neutron emission 
are not strongly dependent on the initial fragment temperature.

\section{Applications}

We have so far shown that there are characteristic correlations 
between the emitted
neutrons that depend on relative angle, energy and multiplicity.  To best
take advantage of these correlations, fast response detector systems are
desirable.  Such systems can better exploit these correlations which would
be washed out in slow response detectors and detection systems based on 
moderators.  Scintillator detectors can distinguish between neutrons and
photons with good background rejection.  They are also amenable to scaling to
larger solid angle coverage.  Threshold detectors that gate on higher-energy
prompt neutron emission can be readily used to study angular-energy or
multiplicity-energy correlations \cite{AdamB}.

As we have shown in Figs.~\ref{nncorr_Pu_fig}-\ref{nncorr_U_fig}, the
neutron-neutron angular correlations can distinguish between configurations
where both neutrons are emitted from a single fragment or one neutron is
emitted from each fragment.  This correlation will become stronger with neutron
energy, particularly for $^{240}$Pu(sf) and $^{238}$U(sf) where the mean
neutron multiplicity is rather low and the emitted spectra are softer.
Such evident directionality could improve background rejection of neutrons
from cosmogenic sources.  Comparison of correlations in admixtures of plutonium
or uranium isotopes could reveal the degree to which the material is enriched.

Figures~\ref{spectra_Pu_fig}-\ref{spectra_U_fig} clearly show the difference
in the spectral shapes for specified neutron multiplicities.  
The slope of the spectral shape increases with neutron multiplicity so,
for example, there are fewer energetic neutrons for $\nu=3$ than for $\nu=1$. 
Again the difference in the spectral distributions gated on neutron 
multiplicity in neutron-induced fission relative to spontaneous fission in
the same sample of material could be exploited by fast detector systems,
providing an additional means of determining the isotopic content of the
material.

For experimental groups to better explore the possible correlation studies
available with \code, we are providing a version to work in-line with several
larger Monte Carlo codes, including $\mathtt{MCNP}$ \cite{MCNP}.  
More details about the
in-line version of \code\ will be discussed in the future.

\section{Concluding remarks}

We have shown that event-by-event models of fission, such as \code,
provide a powerful tool for studying fission neutron correlations.  
Our results demonstrate that these correlations are significant 
and exhibit a dependence on the fissioning nucleus.

Since our method is phenomenological in nature, good input data are especially
important.  Some of the measurements employed in \code\ are rather old and
statistics limited. It would be useful to repeat some of these studies with
modern detector techniques.  
In addition, most experiments made to date have not made simultaneous 
measurements of the fission products and the prompt observables, 
such as neutrons and photons. 
Such data, while obviously more challenging to obtain, would be valauble 
for achieving a more complete understanding of the fission process.

\section*{Acknowledgements}
We wish to acknowledge helpful discussions with 
A.~Bernstein, D.A.~Brown, and C.~Hagmann.
This work was supported by the Office of Nuclear Physics
in the U.S.\ Department of Energy's Office of Science under Contracts No.\ 
DE-AC02-05CH11231 (JR) and
DE-AC52-07NA27344 (RV) 
and by the National Science Foundation, Grant NSF PHY-0555660 (RV).
This research is also supported by the US Department of Energy National 
Nuclear Security Administration Office of Nonproliferation and Verification
Research and Development.

\begin{widetext}

\begin{minipage}{3.35in}
\includegraphics[angle=0,width=\columnwidth]{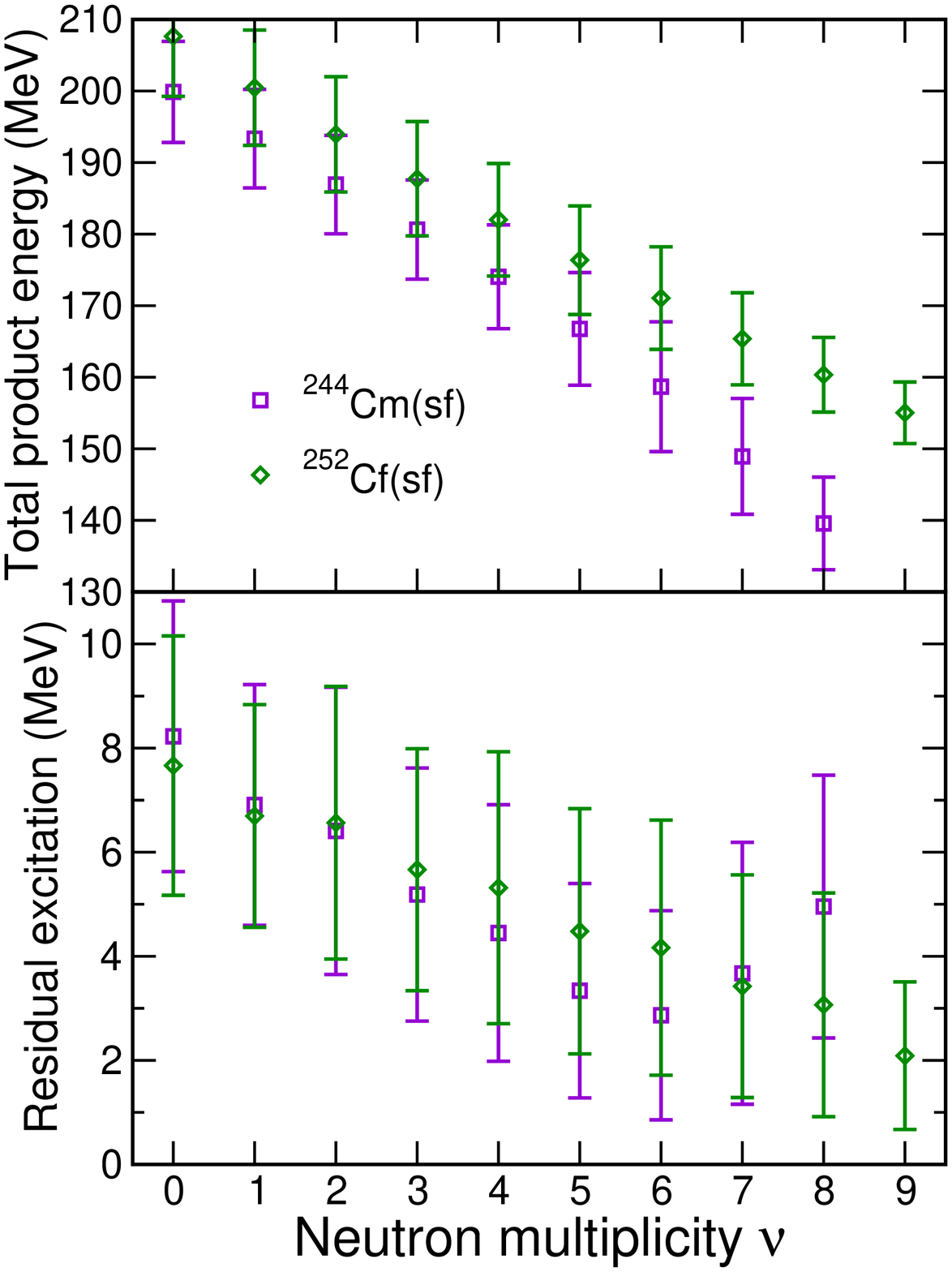}
\phantom{text}\\
{\small FIG.~23: (Color online) The total product kinetic energy (top) and 
residual excitation energy (bottom) remaining after neutron emission has 
ceased as a function of neutron multiplicity.  The \code\ results are shown
for spontaneous fission of $^{244}$Cm (squares) and
$^{252}$Cf (diamonds).}
\end{minipage}
\,\hfill\
\begin{minipage}{3.35in}
\includegraphics[angle=0,width=\columnwidth]{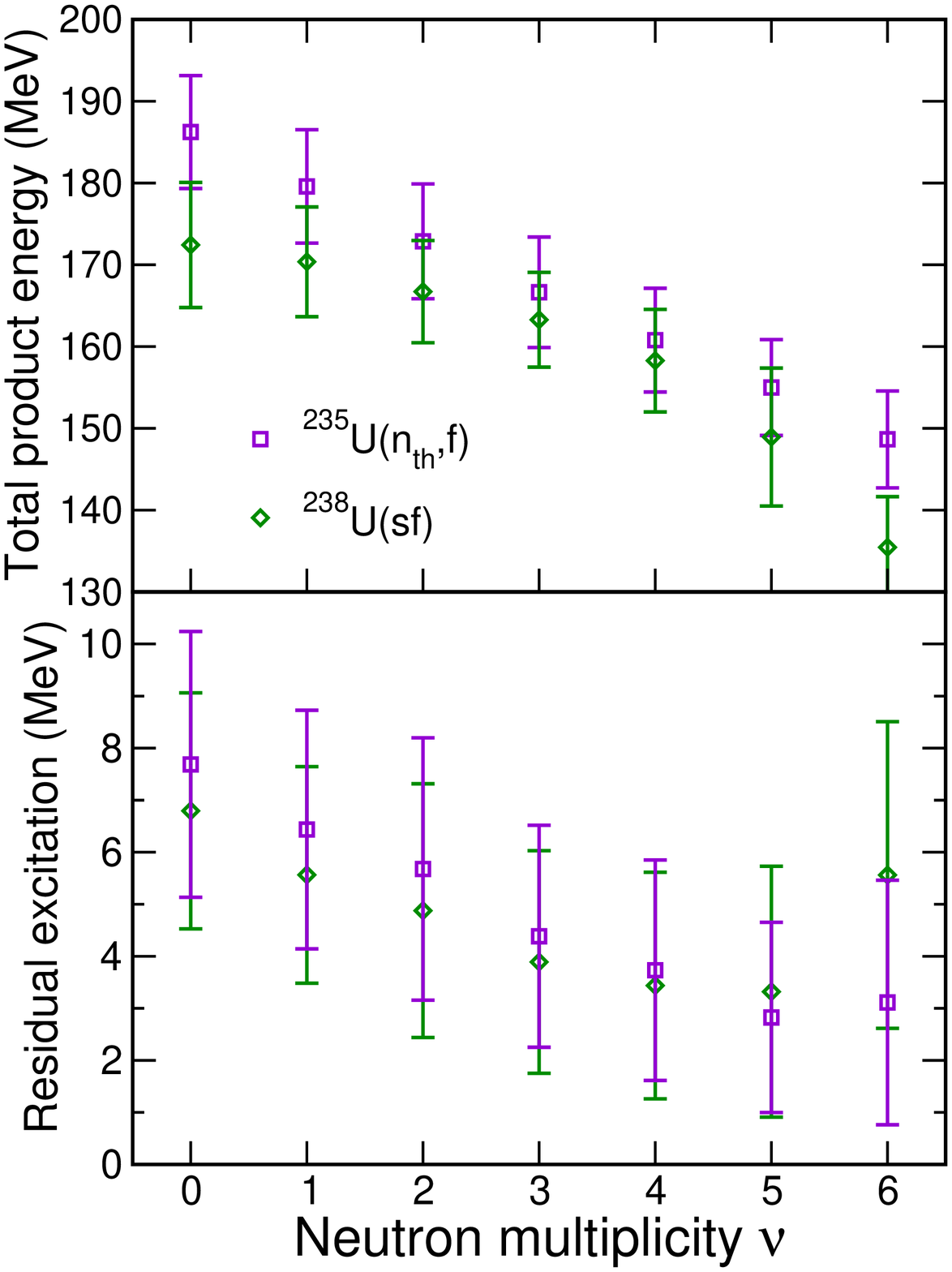}
\phantom{text}\\
{\small FIG.~24: (Color online) The total product kinetic energy (top) and 
residual excitation energy (bottom) remaining after neutron emission has 
ceased as a function of neutron multiplicity.  The \code\ results are shown
for $^{235}$U(n$_{\rm th}$,f) (squares) and $^{238}$U(sf) (diamonds).}
\end{minipage}
\\[2ex]

\newpage
\end{widetext}

\end{document}